\documentclass[11pt]{amsart}

\usepackage{bbm, amssymb,amsmath,graphicx, times,color, multicol, amsthm,url,bigints,mathptmx,enumitem,delarray,verbatim,soul,float,soul,longtable, comment, cancel}
\usepackage[authoryear, round]{natbib}
\usepackage[normalem]{ulem}
\useunder{\uline}{\ul}{}
\usepackage[hidelinks]{hyperref}
\usepackage{graphicx}
\usepackage{caption}
\usepackage{subcaption}

\usepackage{pdflscape}
\usepackage{srcltx}
\usepackage{grffile}
\usepackage[hmargin=0.8in,height=8.8in]{geometry}
\usepackage{mathptmx} 
\usepackage{array}
\usepackage{booktabs}
\usepackage{arydshln}
\graphicspath{{./Images/}}
\newcommand {\diff}{{\rm d}}

\newcommand {\myBox}{\hspace{\stretch{1}}$\diamondsuit$}

\numberwithin{equation}{section}
\newtheorem{proposition}{Proposition}[section]
\newtheorem{remark}{Remark}[section]
\newtheorem{lemma}{Lemma}[section]

\newtheorem{assump}{Assumption}
\newtheorem{corollary}{Corollary}[section]
\newcommand {\R}{\mathbb{R}}

\newcommand {\F}{\mathcal{F}}

\newcommand {\B}{\mathcal{B}}
\newcommand {\p}{\mathbb{P}}
\newcommand {\q}{\mathbb{Q}}

\newcommand {\G}{\mathcal{G}}
\newcommand {\E}{\mathbb{E}}

\newcommand{\conn}{\quad\text{and}\quad}

\newcommand{\1}{\mbox{1}\hspace{-0.25em}\mbox{l}}

\newcommand{\oc}{\lambda^A_\alpha}
\newcommand{\eq}{\mathbf{e_q}}

\newcommand{\nn}{\nonumber}

\usepackage{alphalph}

\title{Loss-Given-Default Modeling by post-last passage time process}
\author[M. Egami]{Masahiko Egami}
\address[M. Egami]{Graduate School of Economics,
Kyoto University, Sakyo-Ku, Kyoto, 606-8501, Japan}
\email{egami.masahiko.8x@kyoto-u.ac.jp}

\thanks{Phone: +81-75-753-3430\\
This version: \today. An earlier version of this paper has been circulated under the title ``A New Approach to Estimating Loss-Given-Default Distribution" or ``Post-last exit time process and its application to loss-given-default distribution''. The author is in part supported by Grant-in-Aid for Scientific Research (C) No.23K01467, Japan Society for the Promotion of Science. }

\begin{document}

\maketitle

\begin{abstract}
We bridge the dichotomy between structural and intensity-based credit models by defining default via the last passage time ($\lambda_\alpha$) of a linear diffusion. Because $\lambda_\alpha$ is not a stopping time, this approach naturally generates ``surprise" defaults within an asset-value framework without artificial filtration enlargements. Mathematically, we establish two general results for transient linear diffusions: we derive explicit Laplace transforms for occupation times and characterize post-passage dynamics. Practically, our two-parameter model calibrates seamlessly to default probabilities and Credit Default Swap (CDS) spreads. We obtain firm-specific Loss-Given-Default (LGD) densities—resolving a long-standing open question—and quantify sentiment divergence between equity and credit markets, yielding a unified valuation framework.
\end{abstract}

\section{Introduction}\label{sec:intro}
This study introduces a novel framework for modeling Loss-Given-Default (LGD) that completely eliminates a priori assumptions regarding its functional form or temporal dynamics. In contemporary credit risk management, Credit Default Swap (CDS) quotes typically rely on a standardized, fixed recovery rate assumption (most commonly a $60\%$ LGD). While computationally convenient, this static convention fails to capture the idiosyncratic, dynamic risk profiles of individual firms and creates a fundamental disconnect between market pricing and underlying financial reality. To address this problem, we make contributions on two interlinked fronts:
\begin{enumerate}
\item [1] \textbf{Mathematical Foundations:} We establish two novel analytical results for general transient linear diffusions. First, we derive an explicit formula for the Laplace transform of the occupation time above a threshold prior to killing, providing a solution to a classical problem in diffusion theory formulated by \citet{karlin-book}. Second, we rigorously characterize the exact dynamics of the post-last passage process. These results provide the robust probabilistic framework required to model a firm's final trajectory toward default.

\item [2] \textbf{Endogenous LGD Extraction:} Leveraging these mathematical foundations, we extract a realistic, firm-specific LGD probability density function directly from observable market parameters (i.e., market-implied Probabilities of Default and CDS spreads). This methodology allows us to bypass the standardized 60\% loss convention while maintaining strict measure-theoretic consistency with prevailing market quotes, yielding a unified pricing framework without introducing additional state variables. By recovering the full LGD density rather than merely its expected value, our approach offers a highly granular and theoretically sound tool for modern credit risk assessment.
\end{enumerate}

The foundational structural approach of \citet{merton1974} and its practical extensions (e.g., Moody’s KMV) remain standard tools for estimating default probabilities. However, structural models notoriously struggle to generate realistic LGD distributions. If default is defined as the first hitting time of the firm value to a deterministic barrier $c$, the asset value at default necessarily equals $c$, reducing LGD to a degenerate constant. On the other hand, reduced-form (intensity-based) models accommodate surprise defaults and flexible recovery assumptions, but they rely on exogenous filtration expansions that sever the explicit link between default events and the firm’s observable economic fundamentals. \textit{Our approach overcomes this long-standing trade-off by operating strictly within the firm's asset value dynamics while ensuring consistency with market-implied credit data.} Crucially, as rigorously established in Section \ref{sec:calibration-in-detail}, we achieve a genuine ``surprise" default—traditionally the exclusive domain of intensity-based models—and derive the true underlying LGD density without resorting to ad-hoc filtration techniques or arbitrary distributional assumptions.

Essential to this unified framework is our novel implementation of the last passage time of the firm-value diffusion. Specifically, let $X=\{X_t, t\ge 0\}$ represent a linear diffusion modeling the underlying asset dynamics, and let $\mathbb{F}=(\mathcal{F}_t)_{t\ge0}$ be the filtration generated by $X$. We focus on the post-last-passage process $(X_t)_{t>\lambda_\alpha}$, where $\lambda_\alpha := \sup \{t \ge 0 : X_t = \alpha\}$ denotes the last passage time to a distressed state $\alpha$. While classical probability theory \citep{meyer1972} characterizes this as a ``newborn" strong Markov process with a distinct transition semigroup upon restarting at $\lambda_\alpha$, we analyze it directly within the continuum of the original diffusion $X$. By preserving the global sample-path trajectory, we seamlessly capture the continuous asset dynamics required to extract the full LGD probability density.

Because $\lambda_\alpha$ depends on the entire future path of $X$, it is inherently unobservable with respect to the forward filtration; that is, $\lambda_\alpha$ is not an $\mathbb{F}$-stopping time for a standard market observer without insider information. Consequently, even though the underlying asset value moves continuously, the default event triggered at or after $\lambda_\alpha$ arrives as an unpredictable shock to external investors. The economic intuition underpinning this mechanism is profound: the mathematical condition that $X_t$ never returns to $\alpha$ after $\lambda_\alpha$ signifies that the firm has irrevocably crossed ``a point of no return", entering a terminal downward trajectory into structural insolvency. However, because $\lambda_\alpha$ is unobservable in real-time, market participants cannot immediately pinpoint the precise moment the firm is fundamentally ruined. This directly mirrors real-world financial distress, where external observers fail to continuously monitor internal operational decay until the ultimate default erupts as a sudden market shock. By exploiting the subtle information structure of last passage times, our model naturally generates realistic, surprise-driven default events while fully anchoring the resulting LGD distribution in the firm's genuine economic reality.

To mathematically formalize the economic intuition discussed above, our framework directly connects the firm's unobservable deterioration to market observables. Specifically, we calibrate the unobservable post-last passage process to the market via the threshold parameter $\alpha$. This calibration serves as a precise mathematical bridge:  it allows us to explicitly derive the endogenous LGD at the time of default. Consequently, we can determine both the true LGD distribution and the observable default intensity in a unified manner.  \textit{The rigorous mathematical formulation of these concepts---including the closed-form derivation of the LGD probability density---is deferred to Section \ref{sec:model-implementation}.}

\subsection{Related literature}
While industry practice often relies on a fixed recovery rate assumption, extensive empirical literature highlights the necessity of time-varying, stochastic LGD models. Empirical studies by \citet{altman2004, altman2005}, \citet{acharya2007}, and \citet{gambetti2019} demonstrate that recovery rates are strongly correlated with default probabilities and are significantly driven by macroeconomic cycles and industry-wide distress. Building on such evidence and related findings by \citet{schlafer2014}, \citet{doshi2018} extracts risk-neutral stochastic recovery dynamics from CDS spreads within an intensity-based framework, confirming that implied recovery rates are inherently time-varying.

In the context of structural and collateralized models, introducing stochastic recovery traditionally requires appending exogenous risk factors. For instance, \citet{yamashita2013} utilizes a correlated stochastic collateral value process, while \citet{cohen2017} extends a structural model by specifying asset and recovery processes as separate, correlated geometric Brownian motions \citep[see also][]{kijima2009}. In stark contrast to these approaches, our framework does not rely on an arbitrary auxiliary recovery process; instead, we generate the LGD distribution purely endogenously from the dynamics of the underlying asset process via the last passage time.

Mathematically, our methodology builds upon the rich probability literature concerning last passage times, time reversal, and related boundary theories \citep{Doob1957, nagasawa, kunita-watanabe1966, salminen1984, rw-1, borodina-salminen, chung-walsh, revuz-yor}. While not a stopping time, the last passage time has been utilized widely in financial applications as discussed in \citet{nikeghbali-platen}. These applications include credit risk \citep{egami-kevkhishvili-reversal}, valuation of defaultable claims \citep{elliott2000, jeanblanc_rutkowski, jeanblanc2009, coculescu2012}, option valuation \citep{profeta2010, cheridito2012}, and insider trading \citep{imkeller2002}.

\bigskip
The remainder of this paper is organized as follows. Section \ref{sec:math} establishes two fundamental mathematical results for general transient linear diffusions: an explicit formula for the distribution of the occupation time above a specific level $\alpha$, and the characterization of the post-last passage time process. Section \ref{sec:model-implementation} details the core valuation framework, deriving explicit formulas for the default time distribution and the endogenous LGD density. Finally, Section \ref{sec:examples} presents focused case studies using empirical CDS data to demonstrate the practical validity of our model, highlighting its capacity to quantify sentiment divergence and uncover potential mispricing in credit markets.

\section{Two  mathematical results concerning last passage time}\label{sec:math}

In this section, we establish two general mathematical results for transient linear diffusions: the first characterizes the pre-$\lambda_\alpha$ regime (specifically, the occupation time prior to killing), and the second describes the dynamics of the post-$\lambda_\alpha$ process. Both results are of independent mathematical interest in their own right, due to their generality, and serve as the foundational analytical tools for our credit risk framework. Because our model (as introduced in the next section) tracks the entire lifetime of the underlying diffusion—representing the firm's asset-to-debt ratio—these two results allow us to analyze the pre- and post-last-passage regimes in a completely unified and consistent manner.

Fix a probability space $(\Omega,\F,\p)$.  Let $X=\{\omega(t),t\ge0;\p_x\}$ be a linear regular canonical diffusion starting at $x\in\R$: that is,
$\omega(t)=X_t(\omega), t\ge 0$ and $\F=(\F_t)_{t\ge 0}$ be a filtration  generated by $X$.
Its state space is given by $\mathcal{I}=(\ell,r)\subset\R$.  Define the shift operator $\{\theta_t, t\ge 0\}$ in the following manner. For each $t$, $\theta_t$ maps $\Omega$ into $\Omega$ such that
\begin{equation*}\label{eq:shift-operator}
\forall t: (X_s\circ \theta_t)(\omega)=X_s(\theta_t\omega)=X_{t+s}(\omega).
\end{equation*}
We refer the reader to Chapter II in \citet{borodina-salminen} for the basic facts regarding linear diffusions. The scale function and the speed measure of $X$ are denoted by $s(\cdot)$ and $m(\cdot)$, respectively. We assume that $s$ and $m$ are absolutely continuous with respect to the Lebesgue measure and have smooth derivatives:
\begin{equation*}\label{eq:m-s}
m(\diff x)=m(x)\diff x \conn s(x)=\int^x s'(y)\diff y,
\end{equation*} where $m$ and $s'$ are continuous and positive.   If $X$ hits $\ell$ or $r$, it is killed and transferred immediately to the cemetery $\Delta\notin \mathcal{I}$. 
We assume that $X$ is \emph{transient}. In this paper, we call a diffusion transient (following \citet{salminen1984}) if
\[
\forall x\in \mathcal{I}, \; \text{$A$ is compact in $\mathcal{B}(\mathcal{I})$: } \p_x(\lambda_A<\infty)=1 \; \textnormal{with} \; \lambda_A=\sup\{t:\omega(t)\in A\}.
\]
The transience in the above sense is equivalent to one or both of the boundaries being attracting; that is, $s(\ell)>-\infty$ and/or $s(r)<+\infty$. 
To obtain concrete results, we set a specific assumption:
\begin{assump}\normalfont\label{assump-s}
\begin{equation*}\label{eq:assumption}
 s(\ell)>-\infty\conn s(r)=+\infty.
\end{equation*}
\end{assump}
\noindent Then, it holds that
\begin{equation*}
\p_x\left(\lim_{t\to\xi}\omega(t)=\ell\right)=1, \quad \forall x\in \mathcal{I}.
\end{equation*}
Hence the process is killed at the lower boundary $\ell$, which is consistent with our asset-to-debt ratio process.

The infinitesimal drift and diffusion parameters are given by $\mu(\cdot)$ and $\sigma(\cdot)$, respectively. We let $\G$ denote the second-order differential operator
\begin{eqnarray*}\label{eq:diff-operator}
\G f(x)=\frac{1}{2}\sigma^2(x)f''(x)+\mu(x)f'(x), \quad x\in \mathcal{I}.
\end{eqnarray*}
For each $t\ge 0$, the transition function is given by $P_t: \mathcal{I}\times \B(\mathcal{I})\mapsto [0,1]$ such that for all $t,s\ge 0$
\begin{equation*}
	\p\left(X_{t+s}\in A\mid \F_s\right)=P_t(X_s,A), \quad \forall A\in\B(\mathcal{I}) \quad \p-a.s.
\end{equation*}
For each $t>0$ and $x\in \mathcal{I}$, $P_t(x,\cdot):A\mapsto P_t(x,A)$ is absolutely continuous \textit{with respect to the speed measure $m$}:
\begin{equation*}
	P_t(x,A)=\int_Ap(t;x,y)m(\diff y), \quad A\in\B(\mathcal{I}).
\end{equation*}
The transition density $p$ may be taken to be jointly continuous in all variables and symmetric such that $p(t;x,y)=p(t;y,x)$.
Note that the diffusion has its \emph{Green function} $G(y, z)$ for $y, z\in \mathcal{I}$ has a representation in terms of its scale function and with respect to its speed measure:
\begin{equation}\label{eq:G0}
  G(y, z):=\int_0^\infty p(t; y, z)\diff t=
  \begin{cases}
   \lim\limits_{a\downarrow\ell, b\uparrow r}\frac{(s(y)-s(a))(s(b)-s(z))}{s(b)-s(a)}, \quad \ell<y\le z <r,\\
\lim\limits_{a\downarrow\ell, b\uparrow r}\frac{(s(z)-s(a))(s(b)-s(y))}{s(b)-s(a)}, \quad \ell <z\le y<r.
  \end{cases}
\end{equation}
The Laplace transform of the hitting time $H_z:=\inf\{t\ge 0:X_t=z\}$ for $z\in \mathcal{I}$ is given by
\begin{equation}\label{eq:hitting-time-laplace}
	\E_x\left[e^{-qH_z}\right]=\begin{cases}
		\frac{\phi_q(x)}{\phi_q(z)}, \quad x\ge z,\\
		\frac{\psi_q(x)}{\psi_q(z)}, \quad x\le z,
	\end{cases}
\end{equation}
where the continuous positive functions $\psi_q$ and $\phi_q$ represent linearly independent solutions of the ODE $\G f=qf$ with $q>0$. Note that $\psi_q$ is increasing while $\phi_q$ is decreasing. These functions are unique up to a multiplicative constant, once the boundary conditions at $\ell$ and $r$ are specified. Finally, the \emph{q-Green function} is defined as
\begin{equation}\label{eq:green-q}
	G_q(x,y):=\begin{cases}
		\frac{\psi_q(y)\phi_q(x)}{w_q}, \quad x\ge y, \\
		\frac{\psi_q(x)\phi_q(y)}{w_q}, \quad x\le y,
	\end{cases}
\end{equation}
with the \emph{Wronskian} $w_q:=\psi_q^+(x)\phi_q(x)-\psi_q(x)\phi_q^+(x)=\psi_q^-(x)\phi_q(x)-\psi_q(x)\phi_q^-(x)$. It holds that $G_q(x,y)=\int_{0}^{\infty}e^{-qt}p(t;x,y)\diff t$ for $x,y\in \mathcal{I}$.

\subsection{Occupation time above $\alpha$ before the last passage time}\label{sec:occupancy}
Our first  result concerns the occupation time of a transient diffusion above a specified threshold $\alpha \in \mathcal{I}=(\ell, r)$ prior to its killing. A related example can be found in the classic textbook by \citet{karlin-book} (Chapter 15, Section 8-B), which considers a diffusion killed upon its first passage to the left boundary and examines the time change driven by the occupation time on the positive half-line. In our framework, the last passage time $\lambda_\alpha$ represents the definitive final instant the process resides in the region above $\alpha$. Building on this structure, we derive the explicit distribution---characterized via its Laplace transform---of the occupation time above $\alpha$ for a general transient diffusion.  Note that a recurrent diffusion can be made transient by designating a finite state as a killing boundary.  To the best of our knowledge, this explicit distributional result has not been previously established in the literature.

Translated to the context of credit risk, this occupation time represents the duration a firm's asset-to-debt ratio remains above the critical threshold $\alpha$. This distribution can be explicitly computed relying solely on the estimated parameters from \eqref{eq:Y/B_2}. Such quantitative metrics offer valuable forward-looking insights for corporate management, lending institutions, and investors. As a straightforward practical application, one can set the killing boundary to $\ell=0.4$ for the process representing asset-to-debt ratio, thereby alinging  the benchmark LGD to the standard market convention of $60\%$.

Fix some $\alpha\in \mathcal{I}$, consider an occupation time of the region above  $\alpha$
\begin{equation*}\label{eq:occup_time}
	\Gamma_+(t):=\int_{0}^{t}\mathbf{1}_{\{X_s\ge\alpha\}}\diff s, \quad t\ge 0,
	\end{equation*}
together with its right inverse:
\begin{equation*}\label{eq:inverse_occup}
	\Gamma_+^{-1}(t):=\inf\{s:\Gamma_+(s)>t\}.
\end{equation*}
Now let us define $\lambda^A_\alpha:=\Gamma_+(\lambda_\alpha)$, which is the \emph{occupation time of $X$ above level $\alpha$ up to the last passage time} $\lambda_\alpha$.  Under Assumption \ref{assump-s}, it is almost sure finite.  We will use the following time-changed process:
\begin{equation*}\label{eq:X-time-change}
	\hat{X}^A(t):=X(\Gamma_+^{-1}(t)).
\end{equation*}
This process $\hat{X}^A$ is illustrated as FIG. 4 in the aforementioned example of \cite{karlin-book} and the letter $\zeta$ there should read $\lambda^A_\alpha$ in the current notation.

Let us introduce another diffusion $X^A$ on $[\alpha, r)$ for which $\alpha$ is a reflecting boundary. The process $X^A$ has the same speed measure and scale function as $\hat{X}^A$ (hence the same  as $X$ as shown in \citet[Theorem 10.12]{dynkin1}) but its killing measure is \emph{zero}. To be precise, we denote
\begin{equation*}\label{eq:Xhat}
	\hat{X}^A(t)=\begin{cases}
		X^A(t), \quad 0\le t <\lambda_\alpha^A\\
		\Delta, \quad t\ge \lambda_\alpha^A
	\end{cases}
\end{equation*} to distinguish the killed process $\hat{X}^A$ from $X^A$.

Before we proceed further, we need to introduce the local time at $\alpha$ for $\hat{X}^A$ by denoting
\begin{equation*}\label{eq:local-time-definition}
	\hat{L}^A(t):=\hat{L}^A(t, \alpha)=L(\Gamma^{-1}_+(t), \alpha)
\end{equation*} where $L(\cdot, \alpha)$ is the local time of $X$ at $\alpha$ which we denote by $L(\cdot)$ for brevity. We note that  $X^A(t)=\hat{X}^A(t)$ and $L^A(t)=\hat{L}^A(t)$ on $[0, \lambda_\alpha^A]$ where $L^A(t)$ is the local time at $\alpha$ for $X^A$. Due to this fact, we can deal with $L^A(\cdot)$ rather than $\hat{L}^A(\cdot)$ in the proof of the next proposition.

Let us also define the inverse local time process  $\rho^A(s):=\inf\{t: L^A(t)>s\}$  for $X^A$ (see \cite[Section 7]{getoor1979}) and also
the q-Green function of $X^A$, denoted by $G_q^A$. The latter is, after some computation, $G_q^A(\alpha, \alpha)=\frac{\phi_q(\alpha)}{-\phi_q^+(\alpha)}$ where $\phi_q$ is in \eqref{eq:hitting-time-laplace} and $\phi_q^+$ is the right derivative with respect to the scale function $s$.

\begin{proposition}\label{prop:occupancy}
  Suppose that the state space of a general transient linear diffusion $X$ is $\mathcal{I}=(\ell, r)$.  Under Assumption \ref{assump-s}, the Laplace transform of $\oc$, the occupation time above level $\alpha$ before the last passage time $\lambda_\alpha$, is
 \begin{equation}\label{eq:la-laplace}
  \E_x\left[e^{-q\oc}\right]=\frac{G^A_q(x, \alpha)}{G^A_q(\alpha, \alpha) + G(\alpha, \alpha)}, \quad q>0, \quad x\ge\alpha,
  \end{equation}
  where $G(\alpha, \alpha)=s(\alpha)-s(\ell)$ (see \eqref{eq:G0}).  Note that $\ell$ can be a finite ($\alpha>\ell$) number or negative infinity.

\end{proposition}

\begin{proof}
 Let $\mathbf{e_q}$ be an exponential random variable, independent of $X$, with rate $q>0$. By the independence of $X$ and $\mathbf{e_q}$, we have
 \begin{align*}
   \E_\alpha\left[e^{-q\oc}\right]=\p_\alpha(\oc< \mathbf{e_q})=\p_\alpha(L^A(\oc)<L^A(\eq)).
 \end{align*}
Since $\lambda^A_\alpha:=\Gamma_+(\lambda_\alpha)$ and $L^A(\oc)=L(\Gamma^{-1}_+(\oc))$, we have
 \[L^A(\oc)=L(\Gamma^{-1}_+(\Gamma_+(\lambda_\alpha))=\lim_{t\rightarrow \lambda_{\alpha}-}L(t)=L(\lambda_\alpha).
 \]
By the argument in Section \ref{sec:excursion}, $L(\lambda_\alpha)<\infty$ is the local time when the first excursion (from $\alpha$) of infinite length occurs and is exponentially distributed, as a Poisson arrival time, with rate $\frac{1}{G(\alpha, \alpha)}$. In other words, corresponding to local time $L(\lambda_\alpha)$, we have a \emph{point} (of the Poisson point process) representing that length of excursion.

On the other hand, it is known that the random variable $L^A(\mathbf{e_q})$ is exponentially distributed and
	\[
	\p_\alpha(L^A(\mathbf{e_q})>s)=\p_\alpha(\rho^A(s)<\mathbf{e_q})=\E_\alpha\left[e^{-q\rho^A(s)}\right]=\exp\left(-\frac{s}{G^A_q(\alpha, \alpha)}\right).
	\]

The random variable $L(\lambda_\alpha)$ is independent of any occasion that corresponds to a certain length of excursion because of the Markov property employed at each time an excursion begins.  In particular, it is independent of $L^A(\eq)$. Therefore, we can compute
\[
\p_\alpha(L^A(\oc)<L^A(\eq))=\frac{\frac{1}{G(\alpha, \alpha)}}{\frac{1}{G(\alpha, \alpha)}+\frac{1}{G^A_q(\alpha, \alpha)}},
\]
which provides \eqref{eq:la-laplace} with $x=\alpha$. For a general $x>\alpha$, by the strong Markov property at $H_\alpha$, we compute
\[
\E_x[e^{-q\lambda_\alpha}]=\E_x[e^{-q(H_\alpha+\lambda_\alpha\circ \theta_{H_\alpha})}]=\E_x[e^{-qH_\alpha}]\cdot \E_\alpha[e^{-q\lambda_\alpha}]
\] and use $\E_x[e^{-qH_\alpha}]=\frac{\phi_q(x)}{\phi_q(\alpha)}$ (see \eqref{eq:hitting-time-laplace}) and \eqref{eq:green-q} to obtain \eqref{eq:occup_time}.  Note that $\phi_q^A(x)=\phi_q(x)$ on $[\alpha,r)$ where $\phi_q^A$ denotes the decreasing solution of $\G^A f=qf$ with the generator $\G^A$ of $X^A$.
\end{proof}
For a broader overview of the excursion-theoretic approach and further details on computing $G_q^A(\cdot, \cdot)$ under various specifications, we refer the reader to Appendices \ref{sec:excursion} and \ref{app:killing-c}, respectively.

\vspace{1em}

\noindent\textbf{Practical Utility for Risk Management:}
Extracting the distribution of the occupation time above the distress threshold $\alpha^*$, as derived here, provides a refined tool for risk management that moves beyond static probability metrics.
\begin{itemize}[leftmargin=*]
    \item Quantifying tail risk: Rather than relying solely on expected values, the full distribution reveals the exact probability of an accelerated structural collapse. Accurately capturing the left tail of this distribution is critical for rigorous stress testing.
    \item Dynamic liquidity and capital optimization: By understanding \textit{how long} a firm is likely to remain in the safe zone, management can formulate time-bound contingency plans and secure additional liquidity on a data-driven timeline.
    \item Precise pricing of credit derivatives: Because crossing the distress threshold $\alpha^*$ initiates a stochastic delay rather than an immediate default, the model generates a smooth, continuous term structure of survival probabilities. This enables highly accurate pricing and hedging of CDS contracts across various maturities.
    \item Covenant breach monitoring: Tracking the distribution of time spent above this critical level serves as a proactive early-warning system for anticipating and managing financial covenant breaches.
\end{itemize}

\subsection{Post-last passage time process (the process conditioned never to return to $\alpha$)}\label{sec:post-last passage}
 In this subsection, we shall identify a general form of the transition density of post-last passage time process in the class of general transient linear diffusions.

We are interested in the distribution of a general transient diffusion $X$ after the last passage time from $\alpha$; i.e., $\p_x(X_t\in \diff y, \lambda_\alpha<t)$.   Recall that for transient diffusion $\lambda_\alpha<\infty$ a.s. Let us define
\begin{equation}\label{eq:h-def}
h_\alpha(x):=\p_x(\lambda_\alpha=0)
\end{equation} and a transform of the transition density of $X$ by
\begin{equation}\label{eq:Meyer-transform}
p^{h_\alpha}(t; x, y):=\frac{h_\alpha(y)}{h_\alpha(x)}p(t; x, y), \quad h_\alpha(x)\neq 0
\end{equation} with respect to the speed measure $m(\diff y)$.

\begin{proposition}\label{prop:semigroup}
	Suppose that the state space of a general transient linear diffusion $X$ is $\mathcal{I}=(\ell, r)$.  Under Assumption \ref{assump-s}, the transition density (with respect to the speed measure $m(\diff y)$) of the post-last passage time process $(X_t)_{t>\lambda_\alpha}$ is given by \eqref{eq:Meyer-transform}.  More precisely,
	\begin{equation}\label{eq:post-density}
		p^{h_\alpha}(t-u; z, y)=
		\frac{h_\alpha(y)}{h_\alpha(z)}p(t-u; z, y), \qquad  y<z<\alpha, \quad u\in (\lambda_\alpha, t)
	\end{equation} for $u$ arbitrarily close to $\lambda_\alpha$: $z$ is arbitrarily close to $\alpha$.
\end{proposition}
\begin{proof}
Let us fix any two points $x,y\in\mathcal{I}\backslash\{\alpha\}$ such that $y<z<\alpha<x$ and consider $\p_x(X_t\in\diff y, \lambda_\alpha<t)$ for an arbitrary $t>0$.  Our objective here is to identify the transition semigroup of the post-last passage time process. For this purpose, we define the first hitting time $H_z:=\inf\{s:X_s=z\}$ as well as $T_z:=\inf\{s>\lambda_{\alpha}: X_s=z\}$. We can assume that when $\lambda_\alpha<t$ holds, $T_z<t$ since $z$ can be set arbitrarily close to $\alpha$.  By Bayes' rule, we have
\begin{align*}
\p_x(X_t\in \diff y, \lambda_\alpha<t)&=\p_x(\lambda_{\alpha}<t\mid X_t\in \diff y)\cdot \p_x(X_t\in\diff y)=\p_x(\lambda_\alpha\circ \theta_t=0 |X_t \in \diff y)\cdot \p_x(X_t \in \diff y, H_z<t) \nn\\
&=\p_y(\lambda_{\alpha}=0)\cdot \E_x\left[\E[\1_{\{X_t\in\diff y\}}\cdot \1_{\{H_z<t\}}\mid \F_{H_z}]\right]\\
		&=h_\alpha(y)\cdot  \E_x\left[p(t-H_z; X_{H_z},y)m(\diff y)\cdot \1_{\{H_z<t\}}\right]\\
&=h_\alpha(y)\cdot  \E_x\left[\left(p(t-H_z; X_{H_z},y)\1_{\{\lambda_\alpha<H_z<t\}}+p(t-H_z; X_{H_z},y) \1_{\{H_z<\lambda_\alpha<t\}}\right)m(\diff y)\right]
\end{align*}
where we used Markov property in the second and strong Markov property together with \eqref{eq:h-def} in the third line.
Let us examine separately the two terms on the right-hand side.  The first term implies that $H_z$ occurs after $\lambda_\alpha$. Hence we use the transform \eqref{eq:Meyer-transform} to obtain
\begin{align*}
 \E_x\left[p(t-H_z; X_{H_z},y)\1_{\{\lambda_\alpha<H_z<t\}}\right]&=\E_x\left[\frac{h_\alpha(X_{H_z})}{h_\alpha(y)}p^{h_\alpha}(t-H_z; X_{H_z}, y)\1_{\{\lambda_\alpha<H_z<t\}} \right]=\E_x\left[\frac{1}{h_\alpha(y)}p^{h_\alpha}(t-T_z; X_{T_z}, y)\1_{\{\lambda_\alpha<H_z<t\}}\right]
\end{align*}
where the last equality is due to $h_\alpha(X_{H_z})=\p_{X_{H_z}}(\lambda_\alpha=0)=1$ and $H_z=T_z$  on the set $\{\lambda_\alpha<H_z<t\}$.

To deal with the second term, we define,
for an arbitrary random time $T<\infty$ a.s.,  a random measure $\tilde{p}(T;u,v)$ such that  $\E_x[\tilde{p}(T;u,v)m(\diff v)]=\p_u(X_{T}\in\diff v)$ for $u,v\in\mathcal{I}$. Note that $\tilde{p}(T_z;x,v)m(\diff v)=0$ a.s. when $v\neq z$. This is because by Fubini's theorem
\begin{align*}
\E_x\left[\int_\mathcal{I}\tilde{p}(T_z;x,v)m(\diff v)\right]&=\int_\mathcal{I}\p_x(X_{T_z}\in\diff v)=\int_\mathcal{I}\int_0^\infty\p_x(X_s\in\diff v, T_z\in\diff s)\\
&=\int_0^\infty\p_x(X_s\in\diff z, T_z\in\diff s)=\p_x(X_{T_z}\in\diff z)=\E_x[\tilde{p}(T_z;x,v)\delta_z(\diff v)]
\end{align*}
and $\p_x(X_{T_z}\in\diff z)=1$ by the definition of $T_z$. Hence the random measure $\tilde{p}(T_z; x, v)m(\diff v)$ has a point mass at $z$. With this preparation the second term becomes
\begin{align*}
\E_x\left[p(t-H_z; X_{H_z},y)\1_{\{H_z<\lambda_\alpha<t\}}\right]
&=\E_x\left[\int_{\mathcal{I}} \tilde{p}(T_z-H_z; z, u)\tilde{p}(t-T_z; u, y)m(\diff u)\1_{\{H_z<\lambda_\alpha<T_z<t\}}\right]\\
&=\E_x\left[p(t-T_z; X_{T_z}, y)\1_{\{H_z<\lambda_\alpha<T_z<t\}}\right]\\
&=\E_x\left[\frac{h_\alpha(X_{T_z})}{h_\alpha(y)}p^{h_\alpha}(t-T_z; X_{T_z}, y)\1_{\{H_z<\lambda_\alpha<T_z<t\}}\right]\nn\\
&=\E_x\left[\frac{1}{h_\alpha(y)}p^{h_\alpha}(t-T_z; X_{T_z}, y)\1_{\{H_z<\lambda_\alpha< T_z<t\}}\right],
\end{align*}
where $h_\alpha(X_{T_z})=1$ is due to the definition of $T_z$. Note that $\1_{\{\lambda_\alpha<H_z<t\}}+\1_{\{H_z<\lambda_\alpha<T_z<t\}}=\1_{\{T_z<t\}}$ by the assumption that  $T_z<t$ whenever $\lambda_\alpha<t$.  Combine these results to obtain
\begin{equation*}\label{eq:before-shift}
  \p_x(X_t\in \diff y, \lambda_\alpha<t)=\E_x\left[p^{h_\alpha}(t-T_z; X_{T_z}, y)\1_{\{T_z<t\}}m(\diff y)\right]=\int_0^t p^{h_\alpha}(t-s; z, y)\p_x(T_z\in\diff s)m(\diff y).
\end{equation*}
Since $z$ is arbitrary, we can make $z$ as close to $\alpha$ as we please.  Therefore, for any time $\lambda_\alpha+u, u>0$, the post- last passage time process is governed by \eqref{eq:post-density} for $u$ arbitrarily close to zero.
\end{proof}

Although  $h_\alpha(x)$ is \emph{not} an excessive function, the transform \eqref{eq:post-density} has  the same form  as if it were an excessive function.  Therefore, we can compute the infinitesimal drift and diffusion parameters of the transform \eqref{eq:Meyer-transform} in the same way as if it is an $h$-transform.  But there is a caveat that $x\neq \alpha$.

Suppose that $X$ is a diffusion (does not have to be transient) and let $h$ be an \emph{excessive} function. Let us denote the infinitesimal drift and diffusion parameters of the $h$-transform by $\mu^h(\cdot)$ and $\sigma^h(\cdot)$, respectively.  The following result is, to our knowledge, new.  See also the argument in Section 15.9 of \citet{karlin-book} that uses a Taylor expansion.
\begin{lemma}\label{lem:drift-diffusion-parameters}
 Assume $h$ is an excessive function and its derivative $h'(y)$ exists for $y\in \mathcal{I}$.  The infinitesimal drift $\mu^h(y)$ and diffusion parameter $\sigma^h(y)$ of the $h$-transform of $Y$ are
  \begin{equation*}\label{eq:hmu-hsigma}
    \mu^h(y)=\mu(y)+\frac{h'(y)}{h(y)}\sigma^2(y) \conn \sigma^h(y)=\sigma(y).
  \end{equation*}
\end{lemma}
\begin{proof}
  See Appendix \ref{app:1}.
\end{proof}

Turning to our case of $h_\alpha$,
\begin{equation}\label{eq:another-view}
h_\alpha(x)=\p_x(\lambda_\alpha=0)=\frac{s(\alpha)-s(x)}{s(\alpha)-s(\ell)}, \quad x\le \alpha.
\end{equation}
We have $\frac{h_\alpha'(x)}{h_\alpha(x)}=-\frac{s'(x)}{s(\alpha)-s(x)}$ so that
\begin{align}\label{eq:new-mu}
	\mu^\alpha(x)=\mu(x)-\dfrac{s'(x)}{s(\alpha)-s(x)}\sigma^2(x),  \conn
	\sigma^\alpha(x)=\sigma(x),
\end{align} where we write $\mu^\alpha$ and $\sigma^\alpha$ instead of $\mu^{h_\alpha}$ and $\sigma^{h_\alpha}$ for simplicity.

Before proceeding further, we should check if the point $\alpha$ is an \emph{entrance boundary} for the post-last passage time process $(X_t)_{t>\lambda_\alpha}$: the process can enter into $(-\infty, \alpha)$ from $\alpha$ but cannot exit from $\alpha$. This confirms that  $(X_t)_{t>\lambda_\alpha}$ cannot return to the region $[\lambda_\alpha, \infty)$.

\begin{proposition}\label{prop:entrance}
  The point $\alpha$ is an entrance boundary for the post-last passage time process $X^{h_\alpha}_t(\omega):=\{\omega(t): \lambda_\alpha<t<H_\ell, \p^{h_\alpha}_\alpha\}$ with $H_\ell := \inf\{t : \omega(t) = \ell\}$.
\end{proposition}
\begin{proof}
See Appendix \ref{app:2}.
\end{proof}

\section{Model Calibration and Computation of LGD distribution}\label{sec:model-implementation}
\subsection{Asset-to-debt ratio process}\label{sec:lev-ratio-prcss}

Operating within the probability space and the general diffusion framework established in Section \ref{sec:math}, we now define the firm's asset-to-debt ratio process, which plays the central role in our credit risk model.  Let us assume that the firm value (market value of total assets) $V$ follows a geometric Brownian motion:
\begin{equation}\label{eq:V-model}
	\diff V_t=\mu V_t\diff t + \sigma V_t\diff W_t
\end{equation}
with $\mu\in \R$ and $\sigma>0$.
The debt process is given by
\begin{equation*}
	B_t=B_0e^{(r_f+r_i) t}, \quad B_0\in\F_0, \quad t\ge 0,
\end{equation*}
where $r_f>0$ denotes a constant risk free rate and $r_i$ is an inflation rate. Modeling debt growth at $r_f + r_i$ accurately reflects real-world corporate financing dynamics through two primary drivers. Baseline interest ($r_f$) represents the continuous, fundamental cost of carrying and rolling over existing debt. On the other hand, in an inflationary environment, operating costs (wages, materials) rise. To maintain the exact same real business scale, a firm requires more cash, forcing it to increase its nominal borrowing proportionally to inflation.

We define the asset-to-debt ratio process $Y=V/B$ on $(0, +\infty)$:
\begin{equation*}
	Y_t=Y_0 e^{\mu-\frac{1}{2}\sigma^2-(r_f+r_i)t+\sigma W_t} \conn  \log\left(Y_t\right)=\log\left(Y_0\right)+\left(\mu-\frac{1}{2}\sigma^2-(r_f+r_i)\right)t+\sigma W_t.
\end{equation*}
The parameters $\mu$, $\sigma$ are such that the left boundary $0$ is attracting and the right boundary $+\infty$ is non-attracting, which ensures that the default occurs eventually with probability one.
In particular, we assume that $\mu-\frac{1}{2}\sigma^2-(r_f+r_i)<0$. By fixing a certain level $\alpha\in \R_+$, we define, as we have done above, \begin{equation*}\label{eq:L-def}
	\lambda_{\alpha}:=\sup\{t\ge 0: Y_t=\alpha\}
\end{equation*}
with $\sup\emptyset=0$. The last passage time is an example of co-optional time satisfying
$\lambda_\alpha\circ \theta_t=(\lambda_\alpha -t)^+, t\ge 0$.
We have the normalized process, with state space $\mathcal{I}=(\ell, r)=(-\infty, +\infty)$,
\begin{equation}\label{eq:Y/B_2}
\frac{1}{\sigma}\diff \log\left(Y_t\right)=\frac{\mu-\frac{1}{2}\sigma^2-(r_f+r_i)}{\sigma}\diff t+\diff W_t=M\diff t+\diff W_t
\end{equation}
by setting
	$M:=\frac{\mu-\frac{1}{2}\sigma^2-(r_f+r_i)}{\sigma}$.
The scale function and the speed measure of this normalized process are 
\[
s(y)=\frac{1}{2M}(1-e^{-2My}) \conn m(\diff y)=2e^{2M y} \diff y.
\]

Now let us introduce the post-last passage time process
\begin{equation} \label{eq:post} Z_t(\omega):=\left(\frac{1}{\sigma}\log(Y_t(\omega))\right)\circ \theta_{\lambda_\alpha(\omega)}(\omega), \quad t>0,
\end{equation} so that $Z$ is subject to the filtration $\F_{\lambda_\alpha + t}$.  \emph{This is the same process as the original $Y$, but  conditioned that it shall never return to $\alpha$.}
The identification of the dynamics of the process $Z$ is undertaken in Section \ref{sec:post-last passage}.

Now we  apply the result in Section \ref{sec:post-last passage} to our asset-to-debt ratio process $Y$ in \eqref{eq:Y/B_2} and its post-last passage time process $Z$ in \eqref{eq:post}.
Using \eqref{eq:new-mu} and the scale function, we see, by a straightforward calculation, that the process $Z_t$ follows the dynamics \eqref{eq:Z-process} right below; that is, the post-$\lambda_{\alpha^*}$ dynamics of the normalized  asset-to-debt ratio process:
\begin{equation}\label{eq:Z-process}
	\diff Z_t=M\coth\left(M\left(Z_t-\alpha^*\right)\right)\diff t + \diff W_t, \qquad Z_0=\alpha^*:=\frac{1}{\sigma}\log(\alpha), \quad t>0,
\end{equation} with $M:=\frac{\mu-\frac{1}{2}\sigma^2-(r_f+r_i)}{\sigma}$.
This is the process we shall work with for explicitly calculating LGD distribution. The starting point of $Z$ is $\alpha^*$. Note that since  $\lim_{x\uparrow 0}\coth(x)=-\infty$, when the process $Z$ approaches $\alpha^*$ from below, the drift approaches negative infinity so that the process shall never reach the point $\alpha^*$ from the region $(-\infty, \alpha^*)$.
Mathematically, as proved in Proposition \ref{prop:entrance}, $\alpha^*$ is an entrance boundary for $Z$ and financially, the firm's asset-to-debt ratio process $Y$ shall never recover to the threshold level $\alpha$.

\subsection{Default time specification}\label{sec:beginning-of-calibration}

The next task is to model the default time $\xi$.  We define a random time that is suitable for default time.
We shall exploit the fact that $\lambda_{\alpha^*}$ is not a stopping time, ensuring that the resulting default time $\xi$ is totally inaccessible: (which is done in Section \ref{subsubsec:filtration_valuation})
\begin{equation}\label{eq:model-xi}
	\xi(\omega):= \lambda_{\alpha^*}(\omega) +\tau\circ \theta_{\lambda_{\alpha^*}(\omega)}(\omega), \quad \xi(\omega)>\lambda_{\alpha^*}(\omega)
\end{equation}
where $\theta(\cdot)$ denotes the shift operator. To specify $\tau$, following \citet[Sec.8.2]{bielecki-rutkowski} and \citet{Filopovic_2009}, we assume that
\begin{assump}\label{assump:technical} There exists a nonnegative $\F_t$- progressively measurable process $\Lambda$ such that $\p[\tau > t|\F_\infty]=e^{-\int_0^t \Lambda(s)\diff s}$ where $\Lambda$ satisfies
\begin{equation}\label{eq:lambda_condition}
	\int_0^\infty \Lambda\left(Z_s\right)\diff s =\infty \quad \p-a.s.
\end{equation}
\end{assump}
This is a standard assumption in the intensity-based credit modeling.  Under Assumption \ref{assump:technical}, proceeding as in Lemma 12.7 of \citet{Filopovic_2009}, we  show that $\p[\int_0^\tau \Lambda(s)\diff s> t |\F_\infty]=e^{-t}$, implying that $\int_0^\tau \Lambda(s)\diff s$ is a unit exponential random variable independent of $\F_\infty$.  We, then,  let $\Lambda: \R_+\mapsto \R_+$ be a nonnegative piecewise continuous function and define $\tau$  by
\begin{equation}\label{eq:moment-inverse}
	\tau=\min\left(s: \int_0^s \Lambda\left(Z_u\right)\diff u=J\right),
\end{equation}
which is the doubly stochastic intensity-based model by \citet{Lando_1998}.  Since $\tau$ is exponential (with rate, say $\eta^p$ under $\p$) and independent of $\lambda_\alpha$, the memoryless property gives
\begin{equation}\label{eq:memoryless}
  \p[\tau \circ \theta_{\lambda_\alpha}> t+s|\tau \circ \theta_{\lambda_\alpha}>s]=\p[\tau \circ \theta_{\lambda_\alpha}>t]=\exp(-\eta^Pt)
\end{equation}
for any $t>0$.  Therefore, \emph{$\tau$ and $\tau\circ\lambda_\alpha$ have the same distribution.}  


It follows from \eqref{eq:moment-inverse} that $\p\left(\tau=0\mid Z_0=\alpha^*\right)=0$ and from \eqref{eq:lambda_condition} that $\p\left(\tau<+\infty\mid Z_0=\alpha^*\right)=1$. We also see that $\p\left(\tau>s\mid Z_0=\alpha^*\right)>0$ for every $s\ge 0$. Moreover, we have   $\p(\lambda_\alpha<\xi<+\infty)=1$ with \eqref{eq:model-xi}.
We have constructed the time $\tau$ using the process which never returns to $\alpha^*$, this feature being represented in \eqref{eq:Z-process}.

Define LGD on debt $B$ as
\begin{equation}\label{eq:KB}
K^B(\xi):=1-Y_\xi=1-\frac{V_\xi}{B_\xi}.
\end{equation}	
Due to $Y_\xi=Y_{\tau}\circ \theta_{\lambda_\alpha}$, it suffices to compute the distribution of $Z_{\tau}=\frac{1}{\sigma}\log(Y_{\tau}\circ \theta_{\lambda_\alpha})$.
Thus, we are able to find the LGD distribution by focusing on $Z_{\tau}$ only. This simplifies the analysis as we do not need to consider the shift operator $\theta$ in the definition \eqref{eq:model-xi}.
Our model is summarized in Table \ref{tbl:setup} and  we have made only Assumptions \ref{assump-s} and \ref{assump:technical}.
\begin{table}[h]
	\caption{ Comparison of our model with the conventional setting}
	\label{tbl:setup}
	\begin{center}
		\begin{tabular}{cccc}
			&Default time & Distribution of the final value\\\hline
			Structural &  $\xi=\inf\left\{t\ge 0:Y_t=c\right\}$ & $Y_\xi=c$ \\ \hline
			Our approach & $\xi=\lambda_\alpha+\tau\circ \theta_{\lambda_\alpha}$ & $Y_\xi\overset{d}{\sim}e^{\sigma Z_{ \tau}}$ \\ \hline
		\end{tabular}
	\end{center}
\end{table}

\subsection{Working under a risk neutral measure}\label{sec:calibration-in-detail}
To calibrate our model to observable market quotes and perform derivative valuation, we transition from the physical measure $\mathbb{P}$ to an equivalent risk-neutral measure $\mathbb{Q}$. Because the asset-to-debt ratio process $Y$ (adapted to $\mathbb{F}$) and the exponential delay $\tau$ are independent under $\mathbb{P}$, the global measure change is defined simply by the product of their respective Radon-Nikodym derivatives.

Let $\eta^{\mathbb{P}}$ and $\eta^{\mathbb{Q}} = \kappa \eta^{\mathbb{P}}$ (with $\kappa > 0$) denote the constant default intensities of $\tau$ under $\mathbb{P}$ and $\mathbb{Q}$, respectively. For a fixed valuation horizon $T$, the Radon-Nikodym derivative $\frac{d\mathbb{Q}}{d\mathbb{P}}$ on the product space is given by
\[
\frac{d\mathbb{Q}}{d\mathbb{P}} = \underbrace{\exp\left(\theta W_T - \frac{1}{2}\theta^2 T\right)}_{\text{Change of drift for } \mathbb{F}} \times \underbrace{\kappa\cdot \exp\left(-(\eta^{\mathbb{Q}} - \eta^{\mathbb{P}})\tau\right)}_{\text{Change of distribution for } \tau},
\]
where $\theta = \frac{r_f - \mu}{\sigma}$ represents the market price of diffusion risk, and $\kappa = \eta^{\mathbb{Q}}/\eta^{\mathbb{P}}$ captures the default-timing risk premium. As we will formally demonstrate in Section \ref{subsubsec:filtration_valuation}, defining the measure change directly on this product space allows us to perform all subsequent valuations entirely within the original asset filtration $\mathbb{F}$, bypassing the conventional requirement for an enlarged filtration $\mathbb{G}$. \emph{Note that importantly by \eqref{eq:memoryless}, the overall default intensity of the model is uniquely specified by the single rate $\eta^{\mathbb{P}}$ under $\mathbb{P}$ and $\eta^{\mathbb{Q}}$ under $\mathbb{Q}$.}

\subsubsection{Estimating $\sigma$}\label{sec:sigma-estimate}
First,we estimate the asset parameters in \eqref{eq:V-model}, in particular $\sigma$, using the maximum likelihood method of \citet{duan1994}, \citet{duan2000}, and \citet{lehar2005}.  Since it is a well-known procedure,  detailed explanation is delegated to Appendix \ref{app:A2}. The current value $Y_0$ of the asset-to-debt ratio process  is retrieved from the last observation in the in-sample data set.

\subsubsection{Calibrating $\alpha^*$}\label{sec:alpha-calibration}
Next, we show the way we  calibrate $\alpha^*$: we make sure that the default time $\xi$ in \eqref{eq:model-xi} matches the market default probability (DP) quoted under the measure $\mathbb{Q}$.  We distinguish between two concepts of default intensity under $\mathbb{Q}$: the market-implied hazard rate under the constant $60\%$ LGD convention, denoted by $\eta^Q_\text{Market}$, and the effective intensity governing the delay $\tau$ in our dynamic LGD model, denoted by $\eta^Q_\text{Model}$.

Under the pure intensity-based framework with a standard constant $60\%$ loss assumption, matching the discounted cash flows of the premium leg and the default leg yields the well-known identity for market CDS spread $S = 0.6 \cdot \eta^Q_\text{Market}$, or equivalently $\eta^Q_\text{Market} = \frac{S}{0.6}$. Intuitively, the instantaneous expected loss upon default, $0.6 \cdot \eta^Q_\text{Market} \diff t$, is continuously compensated by the premium payment $S \diff t$.

However, in our hybrid framework, default is defined via $\xi = \lambda_{\alpha^*} + \tau \circ \theta_{\lambda_{\alpha^*}}$, which inherently implies that no default can occur prior to the last passage time $\lambda_{\alpha^*}$.  To compensate for the zero-default window $[0, \lambda_{\alpha^*})$ and achieve an exact calibration to market quotes, the model intensity $\eta^Q_\text{Model}$ must be scaled relative to the market baseline, so that
  $\eta^Q_\text{Model} = \frac{S}{0.6}(1 + \chi)$ for some $\chi > 0$.
Here, the endogenous multiplicative factor $\chi > 0$ captures the requisite intensity boost during the post-$\lambda_{\alpha^*}$ distress regime, ensuring that our model strictly reconciles with the market-quoted term structure of default probabilities and CDS spreads. Once we fix $\chi$, we can compute the default probability under the model:

 \begin{proposition}\label{prop:calibrate-alpha}
   Under \eqref{eq:model-xi}, the distribution of the default time $\xi$ for each $\chi$  is given by
\begin{small}
\begin{align}\label{eq:5YPD}
\q(\xi\le T)&=\int_0^{T}\left[\left(\int_0^{T-v}\frac{p(u;\frac{1}{\sigma}\log(Y_0),\alpha^*)}{G(\alpha^*,\alpha^*)}\diff u\right)+\left(1-\frac{G(\frac{1}{\sigma}\log(Y_0),\alpha^*)}{G(\alpha^*,\alpha^*)}\right)\right]\eta^Q e^{-\eta^Q v}\diff v \\
&=\frac{e^{M^Q(a-|a|)}}{2M^Q \cdot G(\alpha^*,\alpha^*)}
\int_0^{T}\left(1+\mbox{Erf}\left[\frac{M^Q(T-v)-|a|}{\sqrt{2(T-v)}}\right]-e^{2M^Q|a|}
\mbox{Erfc}\left[\frac{M^Q(T-v)+|a|}{\sqrt{2(T-x)}}\right]\right)\eta^Qe^{-\eta^Qv}\diff v \nn\\ &\hspace{1cm}+\left(1-\frac{G(\frac{1}{\sigma}\log(Y_0),\alpha^*)}{G(\alpha^*,\alpha^*)}\right)(1-e^{-\eta^QT}), \nn
\end{align}
\end{small}
where $\eta^Q=\eta^Q_{\text{Model}}=\frac{S}{0.6}(1 + \chi)$, $p(\cdot, \cdot, \cdot)$ is the transitional density of \eqref{eq:Y/B_2},
\[M^Q=\frac{-\frac{1}{2}\sigma^2-r_{i}}{\sigma}<0, \quad a=\alpha^*-\frac{1}{\sigma}\log(Y_0), \quad
\mbox{Erf}(x)=\frac{2}{\sqrt{\pi}}\int_0^x e^{-t^2}\diff t, \quad  \mbox{Erfc}(x)=1-\mbox{Erf}(x),
\] and
\[
 G(x, y)= \frac{1}{|M^Q|}\exp(M^Q(y-x)-|M^Q||y-x|)
\] with respect to the Lebesgue measure.
 \end{proposition}
\begin{proof}
  See Appendix \ref{app:xi}.
\end{proof}
By equating the model-implied probability on the right-hand side of \eqref{eq:5YPD} to the market quote on the left-hand side, we obtain $\alpha^*$ for each given $\chi$. Importantly, any such solution $\alpha^*$ is unique because the right-hand side of \eqref{eq:5YPD} is strictly increasing with respect to $\alpha^*$. Intuitively, a higher threshold $\alpha^*$ leads to an earlier last passage time $\lambda_{\alpha^*}$, thereby advancing the onset of $\tau$. This strict monotonicity confers a significant computational advantage, ensuring robust and rapid convergence when numerically searching for $\alpha^*$.

\subsubsection{Specifying $\Lambda$}
Following the calibration strategy in Section \ref{sec:alpha-calibration}, a natural specification for the  function $\Lambda$ in \eqref{eq:moment-inverse} under $\mathbb{Q}$ is
\begin{equation}\label{eq:lambda-specific}
	\Lambda(Z_t) := \mathbf{q} \cdot \1_{\{Z_t < \alpha^*\}}, \quad \mathbf{q} > 0.
\end{equation}
Under this formulation, the integral in \eqref{eq:moment-inverse} represents the occupation time of the process $Z$ below the threshold $\alpha^*$. Consequently, $\tau = J / \mathbf{q}$ follows an exponential distribution with rate $\mathbf{q}$, meaning $\mathbf{q}$ serves exactly as the effective model intensity $\eta^Q_{\text{Model}}$. Thus, substituting this into our earlier calibration identity, the hazard rate $\mathbf{q}$ is explicitly given by
\begin{equation}\label{eq:eta_q_scaling}
	\mathbf{q} = \eta^Q_{\text{Model}} = \frac{S}{0.6}(1 + \chi).
\end{equation}

\subsubsection{Determining $\chi$}
To determine the risk premium $\chi$ in \eqref{eq:eta_q_scaling}, we utilize the quoted CDS spread, denoted by $S$.     Suppose that we price the CDS contract based on our default time generated by $\mathbf{q}$ in \eqref{eq:eta_q_scaling}.  Let $D$ and $A$ be the model-implied present value (discounted by $r_f$ under $\q$) of the default leg and premium leg of the contract, respectively.

Before introducing our endogenous LGD dynamics, let us assume a baseline $100\%$ LGD. Under this assumption, the ratio $\left(D/A\right)_\text{Model}(\chi)$, explicitly written as a function of the intensity scaling factor $\chi$, represents the fair annual premium required to compensate for default risk. To ensure the model exactly reconciles with the market CDS spread $S$ (conventionally quoted under a $60\%$ LGD assumption), we equate this model premium to the market-implied hazard rate $\eta^Q_{\text{Market}} = S/0.6$:
\begin{equation}\label{eq:chi-calibration}
	\left(D/A\right)_\text{Model}(\chi) = \frac{S}{0.6}.
\end{equation}
This identity uniquely determines the required parameter $\chi$.

Another important feature of this formulation is the monotonic relationship between $\left(D/A\right)_\text{Model}$ and $\chi$. As shown in \eqref{eq:eta_q_scaling}, an increase in $\chi$ directly leads to a higher hazard rate $\mathbf{q}$. In the computation of $\left(D/A\right)_\text{Model}$, this elevated $\mathbf{q}$ simultaneously increases the numerator (the default leg) and decreases the denominator (the premium leg). Consequently, $\left(D/A\right)_\text{Model}$ is strictly monotonically increasing in $\chi$, guaranteeing a unique root.

Specifically, we determine the pair $(\chi, \alpha^*)$ simultaneously through the following process:

\begin{description}[leftmargin=1.5em, style=unboxed]
    \item[\textbf{Step 1 }] Propose a candidate value for the scaling factor $\chi$.
    \item[\textbf{Step 2 }] For the given $\chi$, numerically find the unique threshold $\alpha^*$ that perfectly aligns the model-implied default probability with the market quote via \eqref{eq:5YPD}.
    \item[\textbf{Step 3 }] Using the proposed $\chi$ and the corresponding $\alpha^*$, compute the model-implied premium $(D/A)_\text{Model}(\chi)$. Evaluate the error against the market target $S/0.6$ as formalized in \eqref{eq:chi-calibration}. Update $\chi$ iteratively (e.g., via Brent's method) until the error converges to zero.
\end{description}
Coupled with the strict monotonicity of the default probability with respect to $\alpha^*$ and the monotonic relationship between $(D/A)_\text{Model}$ and $\chi$, this iterative root-finding process guarantees a unique and computationally efficient solution for the pair $(\alpha^*, \mathbf{q})$. This demonstrates that once appropriately calibrated, our hybrid framework perfectly reproduces market observations without introducing redundant state variables or over-parameterization.

\subsubsection{Market Filtration and Valuation under $\mathbb{F}$}\label{subsubsec:filtration_valuation}
Recall $h_{\alpha^*}(x)=\q_x(\lambda_{\alpha^*}=0)$ as in \eqref{eq:h-def}.  We show that our definition of the default time $\xi$ in \eqref{eq:model-xi} provides particularly convenient mathematical and practical implications.
\begin{lemma}\label{lem:inaccessible}
Let $\mathbb{G} = (\mathcal{G}_t)_{t \ge 0}$ be the market filtration defined by $\mathcal{G}_t := \mathcal{F}_t \vee \sigma(\{\xi \le s\}, \, 0 \le s \le t)$, and let $A^{\alpha^*}_t := \mathbb{Q}(\lambda_{\alpha^*} > t \mid \mathcal{F}_t) = 1 - h_{\alpha^*}(Y_t)$ denote the Az\'ema supermartingale associated with $\lambda_{\alpha^*}$. Then, under $\mathbb{Q}$, the default time $\xi = \lambda_{\alpha^*} + \tau \circ \theta_{\lambda_{\alpha^*}}$ is a $\mathbb{G}$-totally inaccessible stopping time with $\mathbb{G}$-hazard rate
\begin{equation}\label{eq:G_intensity}
\ell^{\mathbb{G}}_t = \mathbf{q} \left( 1 - \frac{A^{\alpha^*}_t}{\mathcal{S}_t} \right), \qquad t \ge 0,
\end{equation}
where $\mathcal{S}_t := \mathbb{Q}(\xi > t \mid \mathcal{F}_t) > 0$ is the $\mathbb{F}$-survival process.
\end{lemma}

\begin{proof}
Since $\{\xi > t\} = \{\lambda_{\alpha^*} > t\} \cup \{\lambda_{\alpha^*} \le t < \xi\}$ and $\tau \circ \theta_{\lambda_{\alpha^*}}$ is exponentially distributed with rate $\mathbf{q}$ independently of $\mathcal{F}_\infty$, conditioning on $\mathcal{F}_t$ yields
\[
\mathcal{S}_t = \mathbb{Q}(\lambda_{\alpha^*} > t \mid \mathcal{F}_t) + \mathbb{E}^{\mathbb{Q}}\!\left[\mathbf{1}_{\{\lambda_{\alpha^*} \le t\}} e^{-\mathbf{q}(t - \lambda_{\alpha^*})} \,\middle|\, \mathcal{F}_t\right] = A^{\alpha^*}_t + \mathbb{E}^{\mathbb{Q}}\!\left[\mathbf{1}_{\{\lambda_{\alpha^*} \le t\}} e^{-\mathbf{q}(t - \lambda_{\alpha^*})} \,\middle|\, \mathcal{F}_t\right].
\]
Note that $\mathcal{S}_t \ge A^{\alpha^*}_t > 0$ for all $t \ge 0$ because $Y$ is a regular transient diffusion on $(0,\infty)$ and $h_{\alpha^*}(y) < 1$ for $y > 0$. By the memoryless property of the exponential distribution, the conditional probability of default occurring in $(t, t + \mathrm{d}t]$ given $\mathcal{F}_t$ satisfies
\[
\mathbb{Q}(t < \xi \le t + \mathrm{d}t \mid \mathcal{F}_t) = \mathbf{q}\, \mathbb{E}^{\mathbb{Q}}\!\left[\mathbf{1}_{\{\lambda_{\alpha^*} \le t\}} e^{-\mathbf{q}(t - \lambda_{\alpha^*})} \,\middle|\, \mathcal{F}_t\right] \mathrm{d}t + o(\mathrm{d}t) = \mathbf{q}(\mathcal{S}_t - A^{\alpha^*}_t)\,\mathrm{d}t + o(\mathrm{d}t).
\]
By standard progressive enlargement of filtration results (e.g., \cite{jeanblanc2009}), the $\mathbb{G}$-compensator of the default indicator process $\mathbf{1}_{\{\xi \le t\}}$ is given by $L_t = \int_0^{t \wedge \xi} \ell^{\mathbb{G}}_s \, \mathrm{d}s$.  Since the hazard rate $\ell^{\mathbb{G}}$ satisfies its defining property: $\mathbb{Q}(t < \xi \le t + \mathrm{d}t \mid \mathcal{F}_t) = \ell^{\mathbb{G}}_t \mathcal{S}_t \, \mathrm{d}t + o(\mathrm{d}t)$, we immediately obtain \eqref{eq:G_intensity}.
Notice that the non-negativity of the hazard rate, $\ell^{\mathbb{G}}_t \ge 0$, is inherently guaranteed by the previously established inequality $\mathcal{S}_t \ge A^{\alpha^*}_t$. This confirms that our intensity process is well-defined and economically meaningful prior to default.
 Since $0 \le \ell^{\mathbb{G}}_t \le \mathbf{q}$, the compensator $L_t$ is continuous, and hence $\xi$ is a $\mathbb{G}$-totally inaccessible stopping time.
\end{proof}

Crucially, although default arrives as a $\mathbb{G}$-totally inaccessible stopping time, all valuation and calibration formulas in our framework can be computed entirely within the original filtration $\mathbb{F}$ without explicitly deriving the $\mathbb{G}$-semimartingale decomposition of $Y$. This is justified by three structural properties: (i) at the valuation time $t=0$, $\mathcal{G}_0 = \mathcal{F}_0$, and for any $0 \le t \le T$ and any bounded $\mathcal{F}_T$-measurable random variable $\Phi_T$, $\mathcal{G}_t$-conditional expectations reduce to $\mathcal{F}_t$-expectations via the standard key lemma,
\begin{equation*}\label{eq:key_lemma}
\mathbb{E}^{\mathbb{Q}}\!\left[\mathbf{1}_{\{\xi > T\}} \Phi_T \,\middle|\, \mathcal{G}_t\right] = \mathbf{1}_{\{\xi > t\}} \frac{\mathbb{E}^{\mathbb{Q}}\!\left[\mathbf{1}_{\{\xi > T\}} \Phi_T \,\middle|\, \mathcal{F}_t\right]}{\mathcal{S}_t},
\end{equation*}
where $\mathcal{S}_t = \mathbb{Q}(\xi > t \mid \mathcal{F}_t) > 0$; (ii) splitting the path at the co-optional time $\lambda_{\alpha^*}$ yields a post-last passage process $Z$ that is itself an autonomous Markov diffusion with explicit transition densities (Proposition~\ref{prop:semigroup}); and (iii) conditional on $\mathcal{F}_\infty$, the post-passage default delay $\tau \circ \theta_{\lambda_{\alpha^*}}$ is exponentially distributed, allowing the default indicator to be integrated out analytically in terms of $\mathbb{F}$-measurable functionals.

\begin{remark}\label{rem:DA_neq_q}\normalfont
Unlike standard reduced-form models with a constant hazard rate, the equality $(D/A)_{\mathrm{Model}} = \mathbf{q}$ no longer holds in our framework. Specifically, under unit LGD, the model-implied default-to-annuity ratio is the survival-weighted time average of the effective $\mathbb{G}$-hazard rate $\ell^{\mathbb{G}}_t$ in \eqref{eq:G_intensity}.  Indeed, since we have
\begin{align*}
  D_{\mathrm{Model}} &:= \int_0^T e^{-r_{f}t} \E^{\mathbb{Q}}[\1_{\{\xi>t\}}\ell^{\mathbb{G}}_t] \diff t
  = \int_0^T e^{-r_{f}t} \E^{\mathbb{Q}}[\mathcal{S}_t\ell^{\mathbb{G}}_t] \diff t
  = \int_0^T e^{-r_{f}t} \E^{\mathbb{Q}}[\mathbf{q}(\mathcal{S}_t-A^{\alpha^*}_t)] \diff t \\
  &= \int_0^T e^{-r_{f}t} \mathbf{q}(\mathbb{Q}(\xi>t) - \mathbb{Q}(\lambda_{\alpha^*}>t)) \diff t
\end{align*}
and similarly $A_{\mathrm{Model}}=\int_0^T e^{-r_{f}t} (\mathbb{Q}(\xi>t) ) \diff t$, we conclude
\[
\left(\frac{D}{A}\right)_{\mathrm{Model}} = \mathbf{q}\left( 1 - \frac{\int_0^T e^{-r_{f}t} \mathbb{Q}(\lambda_{\alpha^*} > t) \, \mathrm{d}t}{\int_0^T e^{-r_{f}t} \mathbb{Q}(\xi > t) \, \mathrm{d}t} \right) < \mathbf{q}.
\]
 Economically, $\mathbf{q}$ represents the true post-passage default intensity once the firm has crossed its point of no return $\lambda_{\alpha^*}$, whereas the observable hazard rate $\ell^{\mathbb{G}}_t$ discounts $\mathbf{q}$ by the conditional probability $\mathbb{Q}(\lambda_{\alpha^*} \le t \mid \xi > t) = 1 - \left(\frac{A^{\alpha^*}_t}{\mathcal{S}_t}\right)$, which is the likelihood that $\lambda_{\alpha^*}$ has already occurred. Because no default can happen on $[0, \lambda_{\alpha^*})$, the ratio $(D/A)_{\mathrm{Model}}$ is strictly smaller than $\mathbf{q}$. In other words, since
\[
\ell^{\mathbb{G}}_t = \mathbf{q} \left( 1 - \frac{A^{\alpha^*}_t}{\mathcal{S}_t} \right) = \mathbf{q} \times \mathbb{Q}(\lambda_{\alpha^*} \leq t \mid \xi > t) + 0 \times \mathbb{Q}(\lambda_{\alpha^*} > t \mid \xi > t),
\]
the intensity $\ell^{\mathbb{G}}_t$ is pushed down by the market's hope that the firm is still in the safe zone ($\lambda_{\alpha^*} > t$). It recovers $\mathbf{q}$ only in the limiting case $\alpha^* \to \infty$ (where $\lambda_{\alpha^*} = 0$ and $A^{\alpha^*}_t = 0$).\myBox
\end{remark}

\subsubsection{Computing LGD}
Finally, we undertake the procedure to compute the distribution of the asset-to-debt ratio at time $\xi$ (i.e., $Y_\xi$), or equivalently that of its post-last passage time process at $\tau$ (i.e., $Z_\tau$) in view of  \eqref{eq:model-xi}.  For this purpose we use the model hazard rate $\mathbf{q}$ in \eqref{eq:lambda-specific}.

As shown in Proposition \ref{prop:entrance}, $\alpha^*$ is an entrance boundary for $Z$. Hence we first consider an arbitrary starting position $z<\alpha^*$ and then take a limit.
Fix any $R<\alpha^*$. According to \eqref{eq:Z-process} and Theorem IV.5.1 in \citet{borodin-book},
\begin{equation}\label{eq:Borodin}
U(z):=\E^\q_{z}\left[e^{-\Lambda\tau}\1_{\{Z_\tau\le R\}}\right], \quad z<\alpha^*, \quad \Lambda>0
\end{equation}
is the unique continuous solution to the ordinary differential equation
\begin{equation}\label{eq:Borodin-solution}
\frac{1}{2}U''(z)+M\coth(M(z-\alpha^*))U'(z)-(\mathbf{q}\Lambda(z)+\gamma)U(z)=-\mathbf{q}\Lambda(z)\1_{[-\infty,R]}(z), \quad z<\alpha^*.
\end{equation}
Here $\E^\q_{z}[\cdot]$ denotes the conditional expectation $\E^\q[\cdot\mid Z_0=z]$. We will provide an explicit form of $U(z)$ in Proposition \ref{prop:joint}.
Using the solution $U(z)$, we obtain $\E^\q_{\alpha^*}\left[e^{-\Lambda\tau}\1_{\{Z_\tau\le R\}}\right]=\lim_{z\uparrow\alpha^*}U(z)$ by the continuity of $U(z)$. Finally, $\lim_{\Lambda\downarrow 0}\E^\q_{\alpha^*}\left[e^{-\gamma\tau}\1_{\{Z_\tau\le R\}}\right]$  produces the distribution of $K^B(\xi)$ via \eqref{eq:KB}.

\begin{proposition}\label{prop:joint}
	Under \eqref{eq:lambda-specific}, fix $\gamma>0$ and let $b_1=\sqrt{1+\frac{2(\mathbf{q}+\gamma)}{(M^Q)^2}}$. For $R<\alpha^*$, we have
	\begin{equation}\label{eq:joint}
		\E^\q_{\alpha^*}\left[e^{-\gamma \tau}\1_{\{Z_{\tau}\le R\}}\right]=\frac{\mathbf{q}}{\mathbf{q}+\gamma}\left(\cosh\left(|M^Q|(\alpha^*-R)\right)+b_1\sinh\left(|M^Q|(\alpha^*-R)\right)\right)e^{-b_1|M^Q|(\alpha^*-R)}.
	\end{equation} 	
\end{proposition}
\begin{proof}
The proofs of this proposition along with the subsequent corollaries are in Appendix \ref{app:U}.
\end{proof}

\begin{corollary}\label{cor:Rtau-dist}
	The distribution of $Z_\tau$ is given by
	\begin{equation}\label{eq:Z-cumdist}
		\q_{\alpha^*}\left(Z_\tau\le R\right)=\begin{cases}
			\left(\cosh\left(|M^Q|(\alpha^*-R)\right)+b_2\sinh\left(|M^Q|(\alpha^*-R)\right)\right)e^{-b_2|M^Q|(\alpha^*-R)}, \quad R\le\alpha^*,\\
			\qquad \qquad \qquad  \qquad \qquad 1, \qquad \qquad \qquad \qquad \qquad \qquad \qquad \qquad  R>\alpha^*,
		\end{cases}
	\end{equation}
	where $b_2=\sqrt{1+\frac{2\mathbf{q}}{M_Q^2}}$.
\end{corollary}

From this result we calculate the distribution of LGD $K^B(\xi)$ in \eqref{eq:KB}:
\begin{corollary}\label{cor:KB-density}
The probability density function of $K^B(\xi)$ in \eqref{eq:KB} is given by
\[
\q_{\alpha^*}(K^B(\xi)\in\diff x) =
\begin{cases}
\frac{2\mathbf{q}}{|M^Q|}\sinh\left(|M^Q|\left(\alpha^*-\frac{\log(1-x)}{\sigma}\right)\right)e^{-b_2|M^Q|\left(\alpha^*-\frac{\log(1-x)}{\sigma}\right)}\frac{1}{\sigma(1-x)}\diff x, & \text{if } 1-e^{\sigma\alpha^*}<x<1 \\
0, & \text{otherwise}
\end{cases}
\]
where $b_2=\sqrt{1+\frac{2\mathbf{q}}{(M^Q)^2}}$. Moreover, the expected value is
\begin{equation}\label{eq:exp-KB}
\E^\q_{\alpha^*}[K^B(\xi)]=1-\frac{\mathbf{q}e^{\sigma\alpha^*}}{\mathbf{q}+\frac{\sigma^2}{2}+b_2\sigma|M^Q|}.
\end{equation}
\end{corollary}

\noindent Therefore, the model-implied spread can be explicitly calculated as follows:
\begin{equation}\label{eq:model-spread}
    \text{Model-Implied Spread} = \text{[Eq. \eqref{eq:exp-KB}]} \times \left(\frac{D}{A}\right)_{\text{Model}}.
\end{equation}

Note that this is a proper probability distribution that integrates to $1$ over the restricted support $(1-e^{\sigma\alpha^*}, 1)$.  The density is strictly zero for $x \le 1 - e^{\sigma\alpha^*}$. Economically, this lower bound represents the model-implied minimum loss given default.

\begin{remark}[Theoretical consistency of the LGD distribution] \normalfont\label{rem:lgd-consistency}
From a theoretical standpoint, extracting the LGD from the distribution of $Z_\tau$ is rigorously justified within our framework. Under the classical option-theoretic valuation of the firm \citep{merton1974}, the exact loss upon default is intrinsically dictated by the firm's residual asset value. Because our hybrid model is exactly calibrated to the market CDS spread, the implied default timing $\xi$ is perfectly consistent with current market pricing. Consequently, mapping this calibrated default mechanism directly to the state of the asset-to-debt ratio $Z$ at time $\tau$ yields an economically sound LGD distribution. It relies solely on the firm's fundamental asset dynamics, completely eliminating the need for arbitrary, exogenous recovery assumptions. \myBox
\end{remark}

\section{Empirical Study}\label{sec:examples}
In this section, we present the empirical application of our model. As established in Section \ref{sec:model-implementation}, our framework relies on exact calibration to market data and derives the LGD distribution endogenously from the  post-last passage time process. As seen in the rows headed ``Relative error'' in Table \ref{tbl:par} below, the calibration to both default probabilities and CDS spreads is highly accurate.   Consequently, this model is not an empirical curve-fitting exercise, nor does it require a large-sample statistical validation to prove its efficacy. Instead, the primary objective of this section is to illustrate the practical mechanics and economic intuition of the model through a focused analysis of a select few representative firms.  By examining these carefully chosen cases, we demonstrate how our structural approach dynamically extracts realistic, firm-specific LGD distributions from market quotes.

\subsection{Data}
We analyze six companies listed on the New York Stock Exchange (NYSE): Delta Air Lines, Inc. (DAL), Ford Motor Company (F), Kohl's Corporation (KSS), Lennar Corporation (LEN), Whirlpool Corporation (WHR), and The Williams Companies, Inc. (WMB). Hereafter, we refer to these firms by their ticker symbols for brevity. Based on the Repository Reporting OTC Data from the Depository Trust \& Clearing Corporation (DTCC), each of these firms recorded an average daily notional amount exceeding 15 million USD in CDS contracts between March 22, 2025, and June 20, 2025. This trading volume indicates sufficient liquidity in the CDS market for our analysis.  For each company, we designate a specific reference date and extract the market-implied LGD distribution using the information available as of that date. Parameter estimation for the dynamics of the asset process $V$  relies on daily data over a 126-business-day trailing window. The constant risk-free rate $r_f$ is proxied by the constant maturity U.S. Treasury yield (investment basis, not seasonally adjusted), which, along with the inflation rate $r_i$, is sourced from the Federal Reserve Economic Data (FRED) database maintained by the Federal Reserve Bank of St. Louis. All the  remaining market data are obtained from LSEG Workspace.

\subsection{Analysis}
We analyze five-year CDS contracts (CDS5) during and following the Covid-19 period. Table \ref{tbl:par} summarizes the empirical and calibration results. The table first reports the estimated parameters of the asset-to-debt ratio process $Y$: the volatility $\sigma$, the drift $M^Q$, and the implied initial state $\frac{1}{\sigma}\log(Y_0)$.

These are followed by the model's fundamental outputs—namely the distress threshold $\alpha^*$, the scaling factor $\chi$, the implied hazard rate $\mathbf{q}$, and the expected LGD $K^B$—derived via the procedure detailed in Section \ref{sec:calibration-in-detail}. Importantly, the results demonstrate the high precision of our calibration framework; the relative errors for both default probabilities and CDS spreads, as formulated in \eqref{eq:chi-calibration}, are remarkably low.

Finally, we compare the market-quoted CDS spreads with the model-implied spreads computed by \eqref{eq:model-spread}, and report the estimated real-world drift $\mu$ of the asset $V$ alongside the theoretical LGD lower bound ($1-e^{\sigma\alpha^*}$) established in Corollary \ref{cor:KB-density}.

\begin{table}[htbp]
\centering
\caption{\scriptsize Estimated parameters up to 4 decimal points.
\textbf{Company industry classifications} in order: Natural Gas Distribution (WMB), Homebuilding (LEN), Consumer Electronics/Appliances (WHR), Auto Manufacturing (F), Air Freight/Delivery Services (DAL), Department/Specialty Retail Stores (KSS).}
\resizebox{\columnwidth}{!}{
\begin{tabular}{lcccccc}
\toprule
 & {\ul \textbf{WMB}} & {\ul \textbf{LEN}} & {\ul \textbf{WHR}} & {\ul \textbf{F}} & {\ul \textbf{DAL}} & {\ul \textbf{KSS}}  \\
\midrule
Evaluation Date                             & 2020-07-02 & 2020-06-03 & 2023-03-30 & 2021-01-13 & 2021-03-26 & 2022-07-07 \\
Risk-free rate  ($r_f$)                     & 0.0029     & 0.0038     & 0.0366     & 0.0048     & 0.0085     & 0.0305     \\
Inflation rate  ($r_i$)                     & 0.0121     & 0.0088     & 0.0240     & 0.0204     & 0.0257     & 0.0253     \\
$\frac{1}{\sigma}\log(Y_0)$             & 1.8326     & 2.2821     & 1.8887     & 1.1566     & 1.6564     & 1.0501     \\
\midrule
$\sigma$                                    & 0.3818     & 0.5034     & 0.3505     & 0.1867     & 0.4347     & 0.5362     \\
$M^Q$                                       & -0.2225    & -0.2691    & -0.2437    & -0.2026    & -0.2764    & -0.3153    \\
$\alpha^*$                                  & -0.8087    & -0.6201    & 0.0816     & -0.0161    & -0.2822    & -0.0751    \\
$\chi$                                      & 68.4931    & 67.9011    & 33.7436    & 45.0904    & 229.5061   & 13.5562    \\
$\mathbf{q}$                                & 0.9190     & 0.9983     & 0.9190     & 1.8781     & 11.5237    & 0.8788     \\
\midrule
Expected value of LGD ($K^B$)               & 0.5549     & 0.6049     & 0.3537     & 0.1712     & 0.2564     & 0.5186     \\
(1) Market PD (\%)                          & 6.522      & 7.241      & 13.715     & 18.726     & 19.680     & 27.779     \\
(1)' Model-implied PD (\%)                  & 6.522      & 7.241      & 13.715     & 18.726     & 19.680     & 27.779     \\
Relative error                              & 0          & 0          & 0          & 0          & 0          & 0          \\
(2) $S/0.6$ (bps)                           & 132.2500   & 144.9000   & 264.5333   & 407.4833   & 499.9333   & 603.7333   \\
(2)' $(D/A)_\text{Model}$ (bps)             & 132.4524   & 147.0421   & 279.9755   & 407.4385   & 499.9333   & 617.6460   \\
Relative error                              & $-1.53 \times 10^{-3}$ & $-1.47 \times 10^{-2}$ & $-5.83 \times 10^{-2}$ & $-1.09 \times 10^{-4}$ & 0 & $2.30 \times 10^{-2}$ \\
\midrule
Market CDS spread                           & 79.35      & 86.94      & 158.72     & 244.49      & 299.96     & 362.24     \\
Effective CDS spread                        & 73.5078    & 88.9460    & 99.0366    & 69.7731   & 128.2030   & 320.3120   \\
$\mu$                                       & -0.1274    & 0.1487     & 0.3227     & 0.0405     & 0.7681     & -0.4873    \\
Minimum boundary                                  & 0.2656     & 0.2681     & -0.0290     & 0.0030     & 0.1154     & 0.0395     \\
\bottomrule
\end{tabular}
}
\label{tbl:par}
\end{table}

\begin{figure}[htbp]
    \centering
    \includegraphics[width=0.8\textwidth]{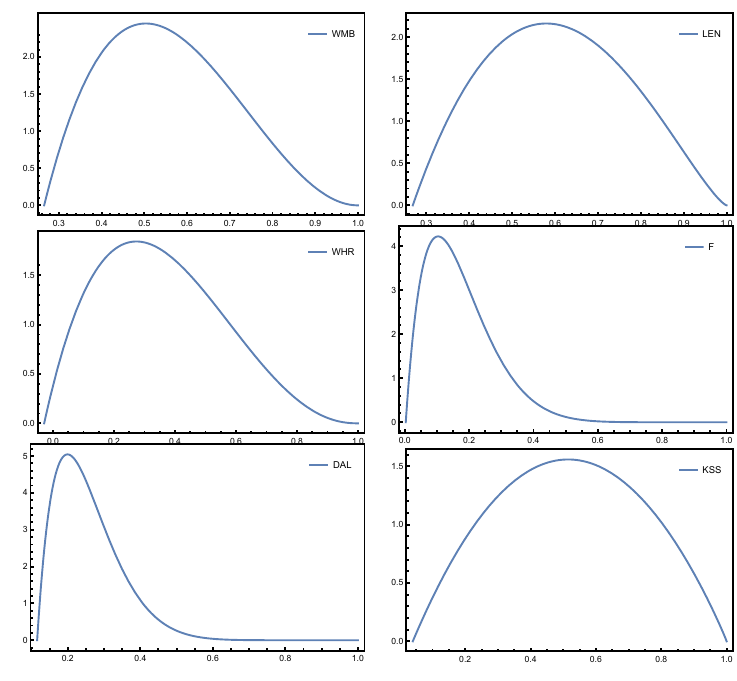}
    \caption{\small{The LGD distributions for the six firms, computed based on the data and calibration results presented in Table \ref{tbl:par}.}}
    \label{fig:LGD}
\end{figure}

\vspace{1em}
\noindent\textbf{Analysis of Loss-Given-Default distribution :} One of the principal theoretical and practical strengths of our proposed framework is its ability to quantify the sentiment divergence between equity and credit markets through the endogenous calibration of the parameter $\chi$. In traditional structural models, a high equity valuation strictly dictates a low probability of default, making it mathematically difficult to reproduce conditions where equity investors are optimistic while the credit market remains pessimistic. Our model overcomes this structural rigidity. For instance, a significantly elevated $\chi$ during the Covid-19 period effectively captures this severe information asymmetry; the equity market prices in strong fundamentals, whereas the CDS market anticipates a sudden default and aggressively prices in a risk premium. By extracting $\chi$ directly from observable data, our framework identifies the possibility of structural mispricing. As demonstrated by Delta Air Lines (DAL), when a firm exhibits a remarkably high $\chi$ but yields an endogenously derived LGD of less than 60\%, it reveals that protection buyers blindly adhering to the market-standard 60\% LGD convention may be significantly overpaying. This capability to translate market sentiment gaps into quantifiable metrics establishes our framework as a highly practical tool for modern credit risk management, achieving this while maintaining the implied hazard rates $\mathbf{q}$ at economically reasonable levels even under extreme stress.

The stability of the hazard rate $\mathbf{q}$ is not merely a numerical artifact; it is a fundamental indicator of the model's structural robustness. In conventional frameworks which lack a parameter like $\chi$, capturing inflated CDS spreads during market panics often forces the calibration algorithm to overcompensate, driving the hazard rate to implausibly high extremes. This overfitting severely distorts the underlying default intensity by conflating pure fundamental risk with sentiment-driven shocks. By allowing $\chi$ to endogenously absorb these market frictions, our framework prevents $\mathbf{q}$ from acting as a catch-all for unexplained risk premiums. Consequently, $\mathbf{q}$ remains reliably anchored to the firm's fundamental credit trajectory. This explicit disentanglement of default risk from market sentiment ensures that the calibrated parameters are economically meaningful, thereby reinforcing the model's theoretical soundness and out-of-sample reliability.

\begin{figure}[htbp]
    \centering
    \includegraphics[width=0.8\textwidth]{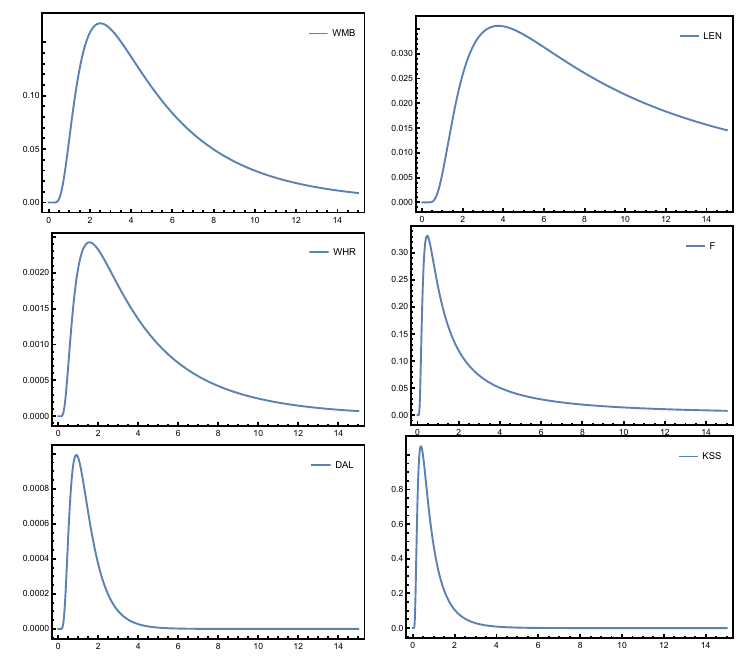}
   \caption{{\small The density functions of the occupation time above $\alpha^*$ before being killed at $c$, computed by the inverse Laplace transform of \eqref{eq:occupancy-concrete}.}}
    \label{fig:Laplace}
\end{figure}

As seen in Figure \ref{fig:LGD}, LGD density shapes vary across firms, yet pairs like F/DAL, LEN/KSS, and WHR/WMB exhibit striking visual similarities. These morphological resemblances are strictly governed by the parameters in Table \ref{tbl:par}.  The expected value $\mathbb{E}^{\mathbb{Q}}_{\alpha^*}[K^B(\xi)]$ anchors the distribution's peak. For instance, F and DAL have notably low expected LGDs, concentrating their density curves in the lower loss regions. Firms sharing a similar expected LGD will naturally have aligned density peaks.
The threshold $\alpha^*$ and asset volatility $\sigma$ define the absolute minimum boundary of the distribution ($1 - e^{\sigma \alpha^*}$). A positive $\alpha^*$ (e.g., WHR) starts the curve near zero, whereas a highly negative $\alpha^*$ (e.g., WMB, LEN) shifts the starting point to a significantly higher minimum loss.  The diffusion parameter $\sigma$ governs the fundamental dispersion of the asset-to-debt ratio. Firms with similar volatility levels exhibit comparable sharpness in their distribution peaks and overall spread.  The hazard rate $\mathbf{q}$ reflects the market's perceived urgency of default and acts as a core exponent in the density function (see Corollary \ref{cor:KB-density}). It directly controls how rapidly the right tail (extreme loss scenarios) decays. Firms with a similar $\mathbf{q}$ price in tail risks at matching rates, resulting in shared distributional skewness and identical asymmetric shapes.

\vspace{1em}
\noindent\textbf{Occupation time over $\alpha^*$:}
Figure \ref{fig:Laplace} displays the probability density function of the occupation time above $\alpha^*$ prior to killing at $c^*<\alpha^*$, denoted by $\lambda_{\alpha^*}^{c^*}$. Here, the normalized process $\frac{1}{\sigma}\log(Y)$ is forcibly killed upon its first passage to the threshold $c^* = \frac{1}{\sigma}\log(c)$. From a risk management perspective, $c^*$ can be interpreted as a critical threshold for strategic decision-making. In our baseline analysis, we set $c^* = 2\alpha^*<0$ for negative $\alpha^*$. Note that for WHR, where $\alpha^*$ is positive, we instead set $c^* = -2$ as a representative lower threshold. The limiting case $c \rightarrow 0$ corresponds to $c^* \rightarrow -\infty$, which represents the lower natural boundary of $\frac{1}{\sigma}\log(Y)$. Refer to Appendix \ref{app:killing-c} for the explicit formula used in this computation.

As demonstrated in the generated distributions, the structural differences in the firms' credit risk (implied by PD and CDS spreads in Table 2) are  captured by the mathematical ``distance'' and ``gravitational pull'' in our model.
\begin{itemize}[leftmargin=*]
    \item Initial buffer (distance to the threshold :  By calculating the distance between the starting asset-to-debt ratio ($\frac{1}{\sigma}\log(Y_0)$) and the distress threshold ($\alpha^*$), the firms can be distinctly categorized into a safe group (WMB, LEN) with substantial buffers ($\frac{1}{\sigma}\log(Y_0) - \alpha^* \approx 2.64$) and a high-risk group (F, KSS) starting very close to the threshold ($\frac{1}{\sigma}\log(Y_0) - \alpha^* \approx 1.13 \sim 1.18$).
    \item Strength of the downward pull (negative drift $M^Q$): Kohl's (KSS), which has the smallest initial buffer, also faces the strongest downward drift ($-0.3153$). It is structurally in the most precarious position.
    \item Link between distributions and market reality: These structural parameters mathematically translate into the inverse Laplace transform graphs. Low-risk firms (WMB, LEN) display distributions that peak further to the right with wider tails, indicating a high probability of remaining above $\alpha^*$. Conversely, high-risk firms (KSS, F) quickly breach the threshold toward the killing boundary ($c^*$), causing their distribution graphs to peak sharply on the far left (near $t=0$), visually quantifying the immediate threat of structural distress.
\end{itemize}

\begin{figure}[h]
	\centering
	\includegraphics[scale=0.75]{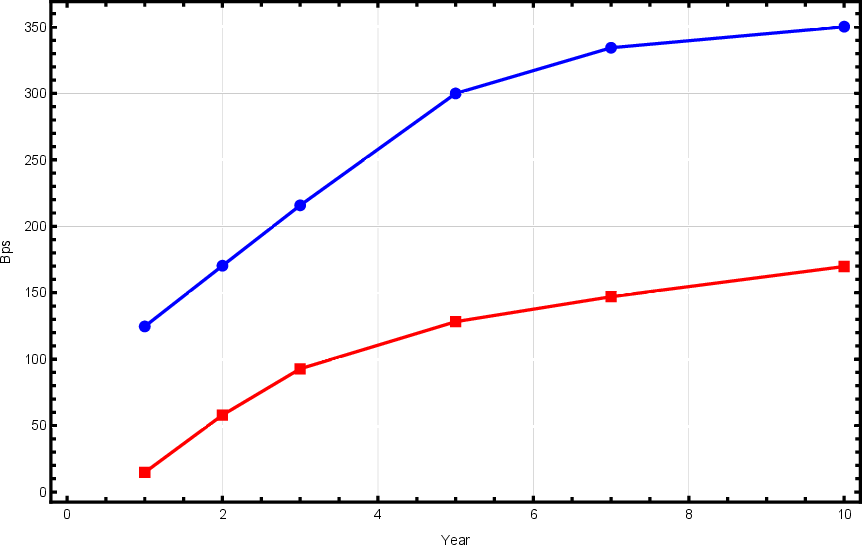}
	\caption{\small{The term structure of DAL CDS contracts (blue: Market, red: Model) on 2021-03-26. The model-implied expected LGDs across different tenors are: 1-year (7.097\%), 2-year (20.409\%), 3-year (25.793\%), 5-year (25.644\%), 7-year (26.375\%), and 10-year (29.076\%).}}
	\label{fig:term-structure}
\end{figure}

\vspace{1em}
\noindent\textbf{Term Structure and Market Implications:} Finally, Figure \ref{fig:term-structure} shows the term structure of Delta Air Lines (DAL) CDS spreads. The red line represents the model-based spreads incorporating the effective LGD, whereas the blue line denotes the market spreads. The structural resemblance between the endogenously derived curve and the empirical market curve validates the model's fundamental premise and its capacity to capture realistic credit risk horizons. Analyzing the gap between these two curves uncovers valuable information regarding market efficiency. Specifically, the market curve exhibits a relatively large jump from the 3-year to the 5-year contract---a feature entirely absent in the model-based curve---which  indicates potential mispricing. By quantifying this divergence, our framework not only offers a more accurate reflection of a firm's credit profile but also highlights arbitrage opportunities for sophisticated market participants seeking to exploit such structural inefficiencies.


\setcounter{section}{0}
\appendix
\section{}
\subsection{Excursion-theoretic (Point process) view }\label{sec:excursion}
We can view $\{X_t, t>\lambda_\alpha\}$ as an excursion from $\alpha$ of \emph{infinite length}.
To pursue this point of view, we review some of the excursion theory based on \citet{getoor1979} and \citet{rw-2}, in particular its Sections VI. 50, 54 and 55.

Let us return to the original process $X$ and consider an excursion from a regular point $\alpha$.  Let the local time at $y\in\mathcal{I}$ be $L(t, y), t\ge 0$. Set $L(t)=L(t, \alpha)$ and  define the right inverse $\rho(t, \alpha):=\inf\{s: L(s)>t\}$.  It is well known that $\rho$ is a right-continuous and nondecreasing \emph{subordinator} under $\p_\alpha$ and hence
\begin{equation}\label{eq:subordinator}
  \E_\alpha\left[e^{-q\rho(t, \alpha)}\right]=e^{-tg(q)}
\end{equation}
where
\begin{equation}\label{eq:exponent1}
g(q)=m\{\alpha\}q+\int_{(0, \infty]}(1-e^{-qs})\nu(\diff s)
\end{equation} with $m\{\alpha\}=0$ in our case and $\nu$ being a measure on $(0, \infty]$ satisfying $\int(1\wedge s)\nu(\diff s)<\infty$.

By using the tail distribution
\[c(u):=\nu((u, \infty]),\]
the Laplace exponent is rewritten as
\begin{equation}\label{eq:exponent2}
	g(q)=m\{\alpha\}q+q\int_0 ^\infty e^{-qs}c(s)\diff s.
\end{equation}

Essentially, $\rho(t, \alpha)$ corresponds to the length of excursion from point $\alpha$ that occurs at local time $t$, so that $\lim_{u\rightarrow \infty}c(u)=\nu(\{+\infty\})$ corresponds to an excursion of infinite length.

We define
\[
r_q(x, y):=\E_x\left(\int_0^\infty e^{-qt}\diff L(t, y)\right), \quad x, y\in \mathcal{I}, q>0.
\] Then, by a change of variable and \eqref{eq:subordinator},
\begin{equation}\label{eq:r_q}
r_q(\alpha, \alpha)=\E_\alpha\left(\int_0^\infty e^{-qt}\diff L(t, y)\right)=\E_\alpha\left(\int_0^\infty e^{-q\rho(t, \alpha)}\diff t\right)=\frac{1}{g(q)}.
\end{equation} But  if we define the $q$-potential $
U^q f(x):=\E_x\left(\int_0^\infty e^{-qt}f(X_t)\diff t\right)$ of a bounded function $f$ with $x\in \mathcal{I}$, as is written in the proof of
Theorem V.50.7 in \citet{rw-2},
\[
U^qf(x)=\int_\mathcal{I} m(\diff y) f(y) \E_x\left(\int_0^\infty e^{-qs}\diff L(s, y)\right)
\] which is also written as
\[U^qf(x)=\int_\mathcal{I} m(\diff y) f(y)G_q(x, y)\]
 by using $G_q(x, y)=\int_0^\infty e^{-qt}p(t; x, y)\diff t$. See \eqref{eq:green-q} for its exact form. By comparison, $G_q(x, y)=\E_x\left(\int_0^\infty e^{-qs}\diff L(s, y)\right)$ a.e. $y$.  But by the right continuity in $y$, we have the equality for every $y\in \mathcal{I}$.
We obtain from \eqref{eq:r_q} that $g(q)=\frac{1}{G_q(\alpha, \alpha)}$ and hence
\begin{equation*}
  \E_\alpha(e^{-q\rho(t, \alpha)})=\exp\left(-\frac{t}{G_q(\alpha, \alpha)}\right).
\end{equation*}
\begin{remark}\normalfont\label{rem:rate}
Let $X$ be a linear diffusion with regular point $\alpha\in \mathcal{I}$.   We consider the local time at $\alpha$ and excursions from this point. With $g(q)=\frac{1}{G_q(\alpha, \alpha)}$, \eqref{eq:exponent2} becomes
  \begin{equation*}\label{eq:laplace-for-u}
 \int_0 ^\infty e^{-qs}c(s)\diff s= \frac{1}{qG_q(\alpha, \alpha)},
  \end{equation*} which can be used to obtain $c(u)=\nu((u, \infty])$ through the inverse Laplace transform. This $c(u)$ corresponds to the rate at which \emph{the first excursion of length greater than $u$} occurs. To consider an excursion of infinite length, which is closely related to a post-last passage time process discussed in Section \ref{sec:post-last passage}, let us take $q\downarrow 0$ in \eqref{eq:subordinator} to find
\[
\p_\alpha(\rho(t, \alpha)<\infty)=e^{-t\nu(\{+\infty\})}
\] by \eqref{eq:exponent1}. Moreover, from $g(q)=\frac{1}{G_q(\alpha, \alpha)}$, we have
\begin{equation*}
	\nu(\{+\infty\})=\lim_{q\downarrow 0}\frac{1}{G_q(\alpha, \alpha)}=\frac{1}{G_0(\alpha, \alpha)}.
\end{equation*}
\myBox
\end{remark}
\bigskip
\subsection{On computations when non-killing (interior) state is made to be killing}\label{app:killing-c}
Let us comment on the killing boundary involved in the computation of \eqref{eq:la-laplace}.  Take an example of (I) Browninan motion with negative drift $M$  (as is the normalized $Y$ process in Section \ref{sec:lev-ratio-prcss}) with two natural boundaries $\ell=-\infty$ and $r=+\infty$, of which $\ell$ is a killing boundary. We, of course, use the Green functions $G(\alpha, \alpha)$ and $G_q^A(x, \alpha)$ for this process to compute the Laplace transform \eqref{eq:la-laplace}. Next, let us consider the case (II) where we set a killing boundary, other than natural boundaries, at some $c\in (\ell, r)=\mathcal{I}$. Then, we should use the q-Green function $G_q^A(x, \alpha)$ for the process ``Brownian motion with drift on $(c, +\infty)$ killed at $c$", which is not the same as that for simple Browinian motion with drift.

For case (I), the increasing and decreasing solutions to $\G f=qf$ are given by $\psi_q(x)=e^{(\sqrt{M^2+2q}-M)x}$ and $\phi_q(x)=e^{-(\sqrt{M^2+2q}+M)x}$. For case (II), let these solutions be denoted by $\psi_q^c$ and $\phi_q^c$ which are linear combinations of $\psi_q$ and $\phi_q$. For the condition at the killing boundary $c$ we refer the reader to \citet[Chapter II.1. Setion 10]{borodina-salminen}:
\[\psi_q^c(c+)=0.\]
We have $\psi_q^c(c)=a_1\psi_q(c)+a_2\phi_q(c)$ from which we obtain

\[a_1=e^{-(\sqrt{M^2+2q}+M)c}, \quad a_2=-e^{(\sqrt{M^2+2q}-M)c}.\]
As there is no boundary condition for $\phi_q^c$, we set $\phi_q^c(x)=\phi_q(x)$. Thus, we have
\begin{align*}
&\psi^c_q(x)=a_1\psi_q(x)+a_2\phi_q(x)=e^{-(\sqrt{M^2+2q}+M)c}e^{(\sqrt{M^2+2q}-M)x}-e^{(\sqrt{M^2+2q}-M)c}e^{-(\sqrt{M^2+2q}+M)x}\\
&\phi_q^c(x)=e^{-(\sqrt{M^2+2q}+M)x}\\
&(\phi_q^c)^+(x)=-(\sqrt{M^2+2q}+M)e^{-(\sqrt{M^2+2q}-M)x}
\end{align*}
together with wronskian $w_q^c=2\sqrt{M^2+2q}\cdot e^{-(\sqrt{M^2+2q}+M)c}$.
Then, $G_q^A(x,\alpha)$ with $x\ge\alpha$ for case (II) is given by
\[\frac{-1}{(\phi_q^c)^+(\alpha)}\phi_q^c(x)\left(\frac{-(\phi_q^c)^+(\alpha)}{w_q^c}\psi_q^c(\alpha)+\frac{(\psi_q^c)^+(\alpha)}{w_q^c}\phi_q^c(\alpha)\right)
=\frac{e^{(\sqrt{M^2+2q}-M)\alpha}\cdot e^{-(\sqrt{M^2+2q}+M)x}}{(\sqrt{M^2+2q}+M)}.\]
Finally, as the introduction of the killing boundary $c$ does not change the scale function, case (II) Green function is given by $G^c(\alpha,\alpha)=s(\alpha)-s(c)$. This is equal to the Green function of the process in case (I) with $\ell$ just being replaced by $c$. This equality is in fact true for every diffusion, not just Brownian motion with drift. For the details of $G^A_q(\cdot, \cdot)$, we refer the reader to \cite{egami-kevkhishvili2025}.

Finally, in drawing the graph of Figure \ref{fig:Laplace} in Section \ref{sec:examples}, we consider the occupation time of the level above $\alpha^*$ before the process $\frac{1}{\sigma}\log(Y)$ hits some level $\frac{1}{\sigma}\log(c)=c^*<\alpha^*$. We denote this time by $\lambda_{\alpha^*}^{A,c}$. Its Laplace transform can be easily calculated using \eqref{eq:la-laplace} and is given
\begin{equation}\label{eq:occupancy-concrete}
  \E_{x^*}\left[e^{-q\lambda_{\alpha^*}^{A,c}}\right]=\frac{\frac{e^{(\sqrt{M^2+2q}-M)\alpha^*}\times e^{-(\sqrt{M^2+2q}+M)x^*}}{w}}
{\frac{e^{(\sqrt{M^2+2q}-M)\alpha^*}\times e^{-(\sqrt{M^2+2q}+M)\alpha^*}}{w}+\frac{1}{2M}(e^{-2Mc^*}-e^{-2M\alpha^*})}
\end{equation}
with $w=\sqrt{M^2+2q}+M$ as derived above.

\bigskip
\subsection{Proof of Lemma \ref{lem:drift-diffusion-parameters} }\label{app:1}
Define $B(x)=\int^x \frac{2}{\sigma^2(y)}\mu(y)\diff y$.  It is well-known that
\begin{equation*}
  m(x)=\frac{2}{\sigma^2(x)}e^{B(x)} \conn s'(x)=e^{-B(x)}
\end{equation*}
and
\begin{equation*}
  m^h(\diff y)=h^2(y)m(\diff y) \conn s^h(\diff y)=\frac{1}{h^2(y)}s(\diff y),
\end{equation*} where $m^h(\cdot)$ and $\sigma^h(\cdot)$ are the speed measure and scale function of the $h$-transform of $X$.  See Section II.1 of \citet{borodina-salminen}.  By simple comparison,  we obtain
\begin{equation}\label{eq:interim}
  \frac{1}{(\sigma^h)^2(x)}e^{B^h(x)}=h^2(x)\frac{1}{\sigma^2(x)}e^{B(x)} \conn e^{-B^h(x)}=\frac{1}{h^2(x)}e^{-B(x)}
\end{equation}
where $B^h(x)=\int^x \frac{2}{(\sigma^h)^2(y)}\mu^h(y)\diff y$.
From \eqref{eq:interim}, we have $\sigma^h(x)=\sigma(x)$.  This in turn gives
\[
\exp\left(2\int^x \frac{\mu^h(y)-\mu(y)}{\sigma^2(y)}\diff y\right)=h^2(x),
\] so that $\mu(x)$ satisfies $\mu^h(x)=\mu(x)+\frac{h'(x)}{h(x)}\sigma^2(x)$.

\bigskip
\subsection{Proof of Proposition \ref{prop:entrance} }\label{app:2}
For the post-last passage time process, we need only to consider a point $y<\alpha$. In view of \eqref{eq:another-view},  \eqref{eq:Meyer-transform} becomes
\begin{equation*}
  p^{h_\alpha}(t; x, y)=\frac{s(\alpha)-s(y)}{s(\alpha)-s(x)}p(t; x, y).
\end{equation*}
This means that the post-last passage time process is obtained equivalently via the transform $g(\cdot):=s(\alpha)-s(\cdot)$. Call this transform also $X^{h_\alpha}$ and its scale function and speed measure $s^{g}$ and $m^g$, respectively. By the general result for the transform of this kind (see Appendix \ref{app:1}), we have
\begin{equation*}
  m^g(\diff y)=g^2(y)m(\diff y) \conn (s^g)'(y)=\frac{1}{g^2(y)}s'(y),
\end{equation*}
which gives us $s^g(y)=\frac{1}{s(\alpha)-s(y)}$. Then, $s^g(\alpha)-s^y(y)=\infty$ for $y<\alpha$ and
\begin{align*}
  \int_y^\alpha(s^g(\eta)-s^g(y))m^g(\diff \eta)
  &=\int_y^\alpha\left(\frac{1}{s(\alpha)-s(\eta)}-\frac{1}{s(\alpha)-s(y)}\right)(s(\alpha)-s(\eta))^2 m(\diff \eta)\\
  &<\int_y^\alpha\frac{1}{s(\alpha)-s(\eta)}(s(\alpha)-s(\eta))^2m(\diff \eta)\\
  &=\int_y^\alpha (s(\alpha)-s(\eta))m(\diff \eta)<\infty.
\end{align*} The last finiteness result is due to $\alpha$ being a regular point for $X$.  This proves that $\alpha$ is an entrance boundary for the $g$-transform, which is equivalent to the post-last passage time process $X^{h_\alpha}$ (see [\cite{karlin-book}, Chap.15, Table 6.2]).

\bigskip
\subsection{Proof of Proposition \ref{prop:calibrate-alpha}}\label{app:xi}
Using \eqref{eq:memoryless} and the fact that $\tau$ is distributed exponentially with rate $\mathbf{q}$ and independent of $\lambda_\alpha$,  we have for $T\in \R_+$
\begin{align*}
  		\q(\xi\le T)&=\q(\lambda_{\alpha^*}\le T-\tau)
  		=\int_0^{T}\q(\lambda_{\alpha^*}\le T-v)\mathbf{q}e^{-\mathbf{q}v}\diff v\\
  		&=\int_0^{T}\q(0<\lambda_{\alpha^*}\le T-v)\mathbf{q}e^{-\mathbf{q}v}\diff v+\int_0^{T}\q(\lambda_{\alpha^*}=0)\mathbf{q}e^{-\mathbf{q}v}\diff v. \nn
\end{align*}

The last passage time distribution is known: Proposition 4 in \citet{salminen1984} and Proposition 3.1 in \citet{egami-kevkhishvili-reversal} provide, in general,
\[
\p_{y}(\lambda_{\alpha}\in \diff u)=\frac{p(u; y, \alpha)}{G(\alpha, \alpha)}\diff u \conn \p_{y}(\lambda_\alpha=0)=1-\frac{G(y, \alpha)}{G(\alpha, \alpha)}\]
where $G(\alpha, \alpha)$ is available in \eqref{eq:G0}. It can be confirmed that $\p_y(\lambda_\alpha=0)=0$ when $y\ge\alpha$ by \eqref{eq:G0} and Assumption \ref{assump-s}.  Since the transition density $p$ for the process $\frac{1}{\sigma}\log(Y_t)$, a Brownian motion with drift $M$, is available, we can evaluate the integral in the above equation explicitly to obtain \eqref{eq:5YPD}.

\bigskip
\subsection{Proof of Proposition \ref{prop:joint}}\label{app:U}
Define $Z^*_t:=\alpha^*-Z_t$. Then, by \eqref{eq:Z-process} under $\p$
\begin{equation*}
	\diff Z^*_t=-\diff Z_t=-M\coth(M(Z_t-\alpha^*))\diff t-\diff W_t=M\coth(MZ^*_t)\diff t+\diff W_t^*=|M|\coth(|M|Z^*_t)\diff t+\diff W_t^*
\end{equation*}
where $(W^*_t)_{t\ge 0}:=(-W_t)_{t\ge 0}$ is a standard Brownian motion. We have used the fact that $-M\coth(-M\cdot x)=	M\coth(M\cdot x)$.

Let us work under $\q$.  We have $\tau=\min\left(s:\mathbf{q}\int_0^s\1_{\{Z_u<\alpha^*\}}\diff u=J\right)=\min\left(s:\mathbf{q}\int_0^s\1_{\{Z_u^*>0\}}\diff u=J\right)$ and
\begin{align}\label{eq:limit}
\E^\q_{\alpha^*}\left[e^{-\gamma \tau}\1_{\{Z_{\tau}\le R\}}\right]&=\lim_{z'\uparrow\alpha^*}\E^\q\left[e^{-\gamma \min\left(s:\mathbf{q}\int_0^s\1_{\{Z_u<\alpha^*\}}\diff u=J\right)}\1_{\{Z_{\tau}\le R\}}\mid Z_0=z'\right] \nn\\
&=\lim_{z\downarrow 0}\E^\q\left[e^{-\gamma \min\left(s:\mathbf{q}\int_0^s\1_{\{Z_u^*>0\}}\diff u=J\right)}\1_{\{Z^*_{\tau}\ge \alpha^*-R\}}\mid Z_0^*=z\right].
\end{align}
By this translation to $Z^*$, \eqref{eq:Borodin} becomes $U(z)=\E^\q\left[e^{-\gamma \min\left(s:\mathbf{q}\int_0^s\1_{\{Z_u^*>0\}}\diff u=J\right)}\1_{\{Z^*_{\tau}\ge \alpha^*-R\}}\mid Z_0^*=z\right]$ and \eqref{eq:Borodin-solution} is now
\begin{align}\label{eq:tau-laplace-odee}
	\frac{1}{2}U''(z)+|M^Q|\coth(|M^Q|\cdot z)U'(z)-\left(\mathbf{q}+\gamma\right)U(z)&=-\mathbf{q}\1_{[ \alpha^*- R,+\infty)}(z), \quad z\in(0,\infty).
\end{align}

\noindent (i) Consider the case when $z\ge \alpha^*-R$. Then \eqref{eq:tau-laplace-odee} is reduced to
\[\frac{1}{2}U''(z)+|M^Q|\coth(|M^Q|\cdot z)U'(z)-\left(\mathbf{q}+\gamma\right)U(z)=-\mathbf{q}.\]
According to Section IV.16.7 and Appendix 2.12 in \citet{borodin-book}, its fundamental solutions are given by
\begin{equation}\label{eq:fund-sol}
	\psi_1(z)=\frac{\sqrt{2}\sinh\left(|M^Q|b_1z\right)}{\sqrt{\pi}b_1\sinh(|M^Q|\cdot z)}, \quad  \phi_1(z)=\frac{\sqrt{\pi}e^{-|M^Q|b_1z}}{\sqrt{2}\sinh(|M^Q|\cdot z)},
\end{equation} where $b_1=\sqrt{1+\frac{2(\mathbf{q}+\gamma)}{(M^Q)^2}}$.
Here $\psi_1$ is an increasing solution and it satisfies $\lim_{z\downarrow0}\psi_1(z)=\sqrt{\frac{2}{\pi}}$ and $\lim_{z\uparrow+\infty}\psi_1(z)=+\infty$ because $b_1>1$. On the other hand, $\phi_1$ is a decreasing solution and it satisfies $\lim_{z\downarrow0}\phi_1(z)=+\infty$ and $\lim_{z\uparrow+\infty}\phi_1(z)=0$. Therefore, for $z\ge \alpha^*- R$, we have $U(z)=\frac{\mathbf{q}}{\mathbf{q}+\gamma}+c\phi_1(x)$ with some constant $c$. We have eliminated the first fundamental solution because $U(z)$ needs to be bounded when $z\to+\infty$.

\noindent (ii) Let $0<z<\alpha^*-R$. In this case, \eqref{eq:tau-laplace-odee} is reduced to
\[\frac{1}{2}U''(z)+|M^Q|\coth(|M^Q|\cdot z)U'(z)-(\mathbf{q}+\gamma)U(z)=0.\]
As in the case (i) above, its fundamental solutions are given by $\psi_1(z)$ and $\phi_1(z)$ in \eqref{eq:fund-sol}.
Therefore, in this case we have  $U(z)=d\psi_1(z)$ with some constant $d$. We have eliminated the second fundamental solution because $U(z)$ is bounded when $z\to 0$.

Finally, from the continuity of $U(z)$ and $U'(z)$ at $\alpha^*-R$, we determine the constants $c$ and $d$.  In particular,

\[d=\frac{\mathbf{q}}{\mathbf{q}+\gamma}\sqrt{\frac{\pi}{2}}\left(\cosh\left(|M^Q|(\alpha^*-R)\right)+b_1\sinh\left(|M^Q|(\alpha^*-R)\right)\right)e^{-b_1|M^Q|(\alpha^*-R)}.\]
Finally, $\lim_{z\downarrow 0}U(z)=\lim_{z\downarrow 0}d\cdot \psi_1(z)=d\cdot\sqrt{\frac{2}{\pi}}$, which proves \eqref{eq:joint} in view of \eqref{eq:limit}.

\bigskip
\subsubsection{\textbf{Proof of Corollary \ref{cor:Rtau-dist} }}
Note that $\q_{\alpha^*}\left(Z_\tau\le R\right)=1$ for $R>\alpha^*$ due to \eqref{eq:Z-process}. For $R\le \alpha^*$, we have
\begin{align*}
	\q_{\alpha^*}\left(Z_\tau\le R\right)=\lim_{\gamma\downarrow 0}	\E^\q_{\alpha^*}\left[e^{-\gamma \tau}\1_{\{Z_{\tau}\le R\}}\right].
\end{align*}
Thus, we obtain the desired result by taking the limit $\gamma\downarrow 0$ in \eqref{eq:joint}.

\bigskip
\subsubsection{\textbf{Proof of Corollary \ref{cor:KB-density}} }
Due to $Y_\xi\overset{d}{\sim}e^{\sigma Z_\tau}$, we have  $K^B(\xi)\overset{d}{\sim} 1-e^{\sigma Z_\tau}$ by \eqref{eq:KB}. The density function of $Z_\tau$ is obtained by differentiating \eqref{eq:Z-cumdist} with respect to $R$:
\begin{equation*}
f(R):=\frac{\q_{\alpha^*}(Z_\tau\in \diff R)}{\diff R}=\frac{2\mathbf{q}}{|M^Q|}\sinh(|M^Q|(\alpha^*-R))e^{-b_2|M^Q|(\alpha^*-R)}\diff R, \quad R<\alpha^*.
\end{equation*}
Then, the density of the transform $1-e^{\sigma Z_\tau}$ is given by
\[\q_{\alpha^*}(K^B(\xi)\in\diff x)=f\left(\frac{\log(1-x)}{\sigma}\right)\frac{1}{\sigma(1-x)}\diff x, \quad 1-e^{\sigma\alpha^*}<x<1.\]
Finally, $\int_{1-e^{\sigma\alpha^*}}^{1}x\q_{\alpha^*}(K^B(\xi)\in\diff x)$ provides \eqref{eq:exp-KB}.

\bigskip
\subsection{Detailed comments for Section \ref{sec:sigma-estimate}}\label{app:A2}
We estimate the parameters of the asset process in \eqref{eq:Y/B_2} using the data of the firm's equity $E$ and debt $B$ available up to the current time. 
	We use the option-theoretic approach of \citet{merton1974} where the value of equity represents the value of the European call option written on the firm's assets with the strike price equal to the future value of debt. We estimate the asset parameters using the maximum likelihood method of \citet{duan1994}, \citet{duan2000}, and \citet{lehar2005}. We assume that debt $B$ grows at the risk-free rate plus inflation $r_f+r_i$. Assuming that debt $B$ is exogenous, the log-likelihood function to be maximized is given by
	\begin{align*}
		&L(E_1,\cdots,E_n; \mu,\sigma\mid B_1,\cdots B_n)=-\frac{n-1}{2}\log(2\pi)-\frac{n-1}{2}\log(\sigma^2\Delta_t)-\sum_{t=2}^{n}\log(\hat{V}_t)-\sum_{t=2}^{n}\log(\Phi(\hat{d}_t))\\
		&-\frac{1}{2\sigma^2\Delta_t}\sum_{t=2}^{n}\left(\log\left(\frac{\hat{V}_t}{\hat{V}_{t-1}}\right)-\left(\mu-\frac{1}{2}\sigma^2\right)\Delta_t\right)^2
	\end{align*}
	where $\Phi$ is a standard normal cumulative distribution function and $\Delta_t$ is the time interval between two consecutive data points. We use $n$ number of past data points for the estimation. In the above equation, $\hat{V}_t$ is the solution to
	\begin{align*}\label{eq:bs}
		E_t=V_t\Phi(d_t(V_t))-B_t e^{r_i T_m}\Phi(d_t(V_t)-\sigma\sqrt{T_m})
	\end{align*}
	with respect to $V_t$ and $\hat{d}_t=d_t(\hat{V}_t)$ where
	\[d_t(V_t)=\frac{\log\left(\frac{V_t}{B_t}\right)+(\frac{\sigma^2}{2}-r_i)T_m}{\sigma\sqrt{T_m}}.\]
	As in \citet{lehar2005}, we set the maturity $T_m$ of the call option to 1 year.
	By differentiating the log-likelihood function with respect to $\mu$ and setting this partial derivative to zero, we obtain
	\[\mu=\frac{\sum_{t=2}^{n}\log\left(\frac{\hat{V}_t}{\hat{V}_{t-1}}\right)}{(n-1)\Delta_t}+\frac{1}{2}\sigma^2.\]
We may plug this into the log-likelihood function which then becomes a function of only one parameter $\sigma$. By maximizing the log-likelihood, we obtain the estimate of $\sigma$ and as a byproduct, the estimate of $\mu$ as well.

\section*{Acknowledgment}
We thank Tomohiro Koike (Graduate School of Economics, Kyoto University) for his valuable comments regarding Proposition \ref{prop:semigroup}.

\end{document}